\documentclass[12pt]{article}

\usepackage[dvips]{graphicx}
\usepackage{epsfig}
\usepackage{amsmath,amsfonts,amssymb,amsthm}
\usepackage{mathrsfs,mathtools}
\usepackage{verbatim}
\usepackage{psfrag}
\usepackage{bm}
\usepackage{bbm}
\usepackage[utf8]{inputenc}
\usepackage[square,comma,sort&compress,numbers]{natbib}
\usepackage[dvipsnames]{xcolor}
\usepackage{slashed}
\usepackage{upgreek}
\usepackage[normalem]{ulem}
\usepackage{enumitem}
\usepackage{float}
\usepackage[T1]{fontenc}
\usepackage{soul}
\usepackage{pgfplots}
\usepackage{extarrows}
\pgfplotsset{compat=1.17}
\usetikzlibrary{external,decorations.pathmorphing}
\usepackage{cases}
\usepackage{listings}
\usepackage{dsfont}

\lstdefinestyle{mystyle}{
    basicstyle=\ttfamily\footnotesize,
    breakatwhitespace=false,         
    breaklines=true,                 
    captionpos=b,                    
    keepspaces=true,                 
    numbers=left,                    
    numbersep=5pt,                  
    showspaces=false,                
    showstringspaces=false,
    showtabs=false,                  
    tabsize=2
}
\usepackage{lmodern}
\usepackage{tikz}

\usepackage[T1]{fontenc}

\tikzset{
  external/only named=true,
  thick/.style={line width=.5pt},
  approximation/.style={line width=1.2pt},
  numerics/.style={black, dotted, line width=.8pt},
  amplitude/.style={dashed},
  estimate/.style={dashed, line width=.8pt},
  normal plot/.style={line width=.8pt},
}

\tikzset{snake it/.style={decorate, decoration=snake}}

\usepackage[colorlinks=true,urlcolor=blue,anchorcolor=blue,citecolor=blue,filecolor=blue
,linkcolor=blue,menucolor=blue,linktocpage=true,pdfproducer=medialab,pdfa=true
]{hyperref}

\usepackage{epsf,epsfig}
\usepackage{graphics}
\usepackage{subcaption}

\newcommand\blfootnote[1]{%
  \begingroup
  \renewcommand\thefootnote{}\footnote{#1}%
  \addtocounter{footnote}{-1}%
  \endgroup
}

\newcommand{\nn}{\nonumber}

\def\d{\mathrm{d}}

\def\C{\mathcal{C}}
\def\O{\mathcal{O}}
\def\vec{\mathbf}

\def\e{\mathrm{e}}
\def\P{\mathcal{P}}

\def\vecp{\vec{p}}

\newcommand{\ket}[1]{\left| #1 \right\rangle}

\usepackage[errorstop]{feynmp}
\newcommand{\lsim}
{\;\raisebox{-.3em}{$\stackrel{\displaystyle <}{\sim}$}\;}

\newcommand{\rom}[1]{\uppercase\expandafter{\romannumeral #1\relax}}

\begin{document}                                                                                                                                                                                                                                                                                                                                                                                                                                                                                                                                                                                                                                                                    

\thispagestyle{empty}

\begin{flushright}
{
\small
CERN-TH-2024-198\\
KCL-PH-TH/2024-57
}
\end{flushright}

\vspace{-0.5cm}

\begin{center}
\Large\bf\boldmath
Bounds on the bubble wall velocity
\unboldmath
\end{center}

\vspace{-0.2cm}

\begin{center}
Wen-Yuan Ai,$^{*1}$\blfootnote{$^*$wenyuan.ai@kcl.ac.uk} Benoit Laurent$^{\dagger 2}$\blfootnote{$^\dagger$benoit.laurent@mail.mcgill.ca} and Jorinde van de Vis$^{\ddagger 3}$\blfootnote{$^\ddagger$jorinde.van.de.vis@cern.ch} \\
\vskip0.4cm

{\it $^1$Theoretical Particle Physics and Cosmology, King’s College London,\\ Strand, London WC2R 2LS, United Kingdom} \\
{\it $^2$McGill University, Department of Physics, 3600 University St.,\\
Montr\'{e}al, QC H3A2T8 Canada} \\
{\it $^3$Theoretical Physics Department, CERN, \\ 1 Esplanade des Particules, CH-1211 Geneva 23, Switzerland}
\vskip1.cm
\end{center}

\begin{abstract}

Determining the bubble wall velocity in first-order phase transitions is a challenging task, requiring the solution of (coupled) equations of motion for the scalar field and Boltzmann equations for the particles in the plasma. The collision terms appearing in the Boltzmann equation present a prominent source of uncertainty as they are often known only at leading log accuracy. In this paper, we derive upper and lower bounds on the wall velocity, corresponding to the local thermal equilibrium and ballistic limits. These bounds are completely independent of the collision terms.

For the ballistic approximation, we argue that the inhomogeneous plasma temperature and velocity distributions across the bubble wall should be taken into account.
This way, the hydrodynamic obstruction previously observed in local thermal equilibrium is also present for the ballistic approximation. This is essential for the ballistic approximation to provide a lower bound on the wall velocity. We use a model-independent approach to study the behaviour of the limiting wall velocities as a function of a few generic parameters, and we test our developments in the singlet extended Standard Model.

\end{abstract}

\newpage

\hrule
\tableofcontents
\vskip .85cm
\hrule


\section{Introduction}
\label{sec:Intro}

First-order phase transitions (FOPTs) are common phenomena in daily life but are also naturally predicted to have occurred in the early Universe by many models beyond the Standard Model (SM). See, e.g., Refs.~\cite{Cline:1996mga,Carena:1997ki,Grojean:2004xa,Fromme:2006cm,Grojean:2006bp,Profumo:2007wc,Barger:2007im,Delaunay:2007wb,FileviezPerez:2008bj,Gil:2012ya,Dorsch:2013wja,Huang:2016odd,Jinno:2016knw,Chao:2017vrq,Beniwal:2017eik,Marzola:2017jzl,Kurup:2017dzf,Chen:2017cyc,Baldes:2018emh,Prokopec:2018tnq,Bian:2018bxr,Marzo:2018nov,Chala:2018opy,Zhou:2018zli,Alves:2018jsw,Azatov:2019png,DelleRose:2019pgi,VonHarling:2019rgb,Halverson:2020xpg,Ghosh:2020ipy,Huang:2020crf,DiBari:2021dri,Kierkla:2022odc,Morgante:2022zvc,Fujikura:2023fbi,Frandsen:2023vhu,Pasechnik:2023hwv,Feng:2024pab,Gao:2024pqm,Gao:2024fhm} for a far from complete list of examples.
FOPTs are interesting not only because they can generate a stochastic background 
of gravitational waves (GWs)~\cite{Witten:1984rs,Hogan:1986dsh,Kosowsky:1992vn,Kosowsky:1992rz,Kamionkowski:1993fg}, but also because they have far-reaching phenomenological applications. Among other possibilities, they may be relevant for explaining the cosmic matter-antimatter asymmetry~\cite{Kuzmin:1985mm,Morrissey:2012db,Garbrecht:2018mrp,Cline:2020jre,Azatov:2021irb,Baldes:2021vyz,Huang:2022vkf,Chun:2023ezg,Cataldi:2024pgt} or the generation of dark matter~\cite{Falkowski:2012fb,Baker:2019ndr,Chway:2019kft,Chao:2020adk,Azatov:2021ifm,Azatov:2022tii,Baldes:2022oev,Giudice:2024tcp,Azatov:2024crd,Ai:2024ikj}.

Almost all the FOPT-related phenomena, as well as the generated GW signals,
depend crucially on the bubble wall velocity $v_w$. For this reason, there have been a lot of works on bubble wall dynamics in recent years~\cite{BarrosoMancha:2020fay,Baldes:2020kam,Friedlander:2020tnq,Balaji:2020yrx,Cline:2021iff,Bea:2021zsu,Bigazzi:2021ucw,Ai:2021kak,Lewicki:2021pgr,Dorsch:2021nje,DeCurtis:2022hlx,Laurent:2022jrs,Wang:2022txy,Lewicki:2022nba,Janik:2022wsx,Li:2023xto,Ai:2023see,Krajewski:2023clt,Giombi:2023jqq,Wang:2023kux,Gouttenoire:2023roe,Dorsch:2023tss,Sanchez-Garitaonandia:2023zqz,DeCurtis:2024hvh,Kang:2024xqk,Wang:2024wcs,Evans:2024ilx,Yuwen:2024hme, Ekstedt:2024fyq}. To determine $v_w$, one in principle needs to solve the coupled equation of motion (EoM) of the order-parameter scalar field and the Boltzmann equations for the particles that couple with the scalar field~\cite{Moore:1995si,Moore:1995ua}. See Refs.~\cite{Liu:1992tn,Dorsch:2018pat,Wang:2020zlf,Laurent:2020gpg,Laurent:2022jrs,Jiang:2022btc, Ekstedt:2024fyq} for some studies following this approach. 
In~\cite{Ignatius:1993qn,Heckler:1994uu,Kurki-Suonio:1996gkq,Espinosa:2010hh,Huber:2011aa,Huber:2013kj}
the backreaction
of the fluid onto the expanding bubbles is incorporated by adding an effective friction term to the scalar field EoM. 

Solving the Boltzmann equations and the scalar EoM is however challenging not only because they are coupled integrodifferential equations, 
but also because the computation of the collision terms suffers from large uncertainties. It is therefore highly desirable to derive the upper and lower bounds of the wall velocity with simpler approaches, which is the goal of the present work. 
Such upper and lower bounds can be very useful in the initial exploration of a BSM model. For example, certain dark matter generation mechanisms require bubbles with very slow walls \cite{Baker:2019ndr, Chway:2019kft}. To find out whether a specific BSM realization can give successful dark matter generation, one might first determine the upper and lower bound for the wall velocity. These results can then motivate a more detailed computation of $v_w$, including a solution of the Boltzmann equations (or save one from wasting time and numerical resources).

Our estimates for the upper and lower bound of the wall velocity correspond to the two limiting cases for the deviation from equilibrium of the particles in the plasma. Intuitively, the larger the deviation from thermal equilibrium of the plasma, the greater the friction the bubble wall experiences. One can imagine (and we will demonstrate in this work) that the deviation from equilibrium is a monotonically decreasing function of the collision rate $\Gamma$ for the particles in the plasma. Let us assume that the interaction strength is infinitely strong, i.e., $\Gamma\rightarrow\infty$, particles can instantaneously relax to their equilibrium state after being perturbed, and therefore, the plasma would remain in local thermal equilibrium as the wall passes through. This is the local thermal equilibrium (LTE) approximation~\cite{Balaji:2020yrx,Ai:2021kak,Ai:2023see}, which can provide the lower bound of the friction experienced by the wall and the upper bound of the velocity. Although in LTE there is no {\it dissipative} friction, there is still a backreaction force due to hydrodynamic effects~\cite{Ignatius:1993qn,Konstandin:2010dm}, which could lead to a stationary motion of the wall. Such hydrodynamic effects are caused by the inhomogeneous temperature and velocity distributions near the wall. It was noted in Ref.~\cite{Ai:2021kak} that in LTE, the wall velocity can be determined very simply and efficiently with a new matching condition in the hydrodynamic quantities across the bubble wall due to the conservation of entropy. This method has been more thoroughly studied in Ref.~\cite{Ai:2023see}. 

In the opposite limit, $\Gamma\rightarrow 0$, particles do not collide with each other, and the friction can be studied by analyzing particle transmission across the wall. This is the {\it ballistic} approximation that is usually used for ultra-relativistic bubble walls~\cite{Bodeker:2009qy,Bodeker:2017cim,Hoche:2020ysm,Azatov:2020ufh,Gouttenoire:2021kjv,GarciaGarcia:2022yqb,Ai:2023suz}, but has also been applied for slow walls in  Refs.~\cite{Liu:1992tn,BarrosoMancha:2020fay,Wang:2024wcs}. In contrast to the LTE case, the inhomogeneity in the plasma temperature and velocity distributions is usually not considered in the ballistic approximation.\footnote{An exception is the recent work~\cite{Wang:2024wcs} which considers the inhomogeneous distribution in the temperature but not in the fluid velocity.} Without considering it, the crucial hydrodynamic effects are ignored and the estimate of the velocity could be misleading. In particular, it has been observed when solving the scalar EoM and Boltzmann equations~\cite{Cline:2021iff,Laurent:2022jrs} (see also Refs.~\cite{DeCurtis:2022hlx,DeCurtis:2023hil}), but also in a pure LTE analysis~\cite{Ai:2023see,Ai:2024shx} that the frictional pressure has a peak at the Jouguet velocity.
Such a peak is important as it tells us that the friction is not a monotonous function of
the wall velocity and thus one cannot simply use the asymptotic value of the friction in the ultrarelativistic limit~\cite{Bodeker:2009qy} to determine whether a bubble wall runs away or not~\cite{Ai:2024shx}. In this work, we carefully account for the mentioned inhomogeneity in the plasma temperature and velocity across the wall. As a result, we observe a similar pressure curve peaked at the Jouguet velocity for the ballistic approximation, and we can extract a lower bound on the wall velocity.

The goal of this work is three-fold:
\begin{itemize}
    \item We update the ballistic approximation to account for the inhomogeneous temperature and velocity profile, promoting it such that it gives us a lower bound on $v_w$.
    \item We demonstrate that the ballistic approximation gives the upper bound of the friction  (equivalently lower bound of the wall velocity), and the LTE approximation the lower bound.
    \item We demonstrate in practice how the bounds on the velocity can be applied to simple and realistic models.
\end{itemize}

In this paper, we will see a few different length scales for the phase transition (PT) dynamics. Here we summarise them in Table~\ref{tab:scales}. The reader may refer to it when these length scales are introduced.

\begin{table}[ht]
    \centering
    \begin{tabular}{|c||c|c|}
        \hline
        Length scale & Description & Typical value \\
        \hline\hline
        \begin{tabular}{@{}c@{}}
        Wall thickness $L_w$\\[-1.7mm] (wall frame) 
        \end{tabular}
          & Scalar fields' variation length scale & $1/m_\phi$ \\
        \hline
        \begin{tabular}{@{}c@{}}
         Mean free path $L_{\rm MFP}$ \\[-1.7mm] (plasma frame)   
        \end{tabular}
        & Plasma's thermalisation length scale & $1/\Gamma\sim 1/(g^n T)$ \\
        \hline
        \begin{tabular}{@{}c@{}}
         Bubble radius $R_{\rm bubble}$ \\[-1.7mm] (plasma frame) 
        \end{tabular}
         & \begin{tabular}{@{}c@{}}Length scale of the macroscopic structures \\[-1.7mm] (e.g.\ shock and rarefaction waves, bubble)\end{tabular} & $\mathcal O(1 - 10^{-5}) \times 1/H$ \\
        \hline
        \begin{tabular}{@{}c@{}}
             $\delta$  \\
             (wall frame) 
        \end{tabular}
         & \begin{tabular}{@{}c@{}}Distance from the wall at which \\[-1.7mm] the scalar fields are constant\end{tabular}  & \begin{tabular}{@{}c@{}} $\delta\gg L_w$ \\[-1.5mm] and $\delta\ll \gamma_w R_{\rm bubble}$\end{tabular}  \\
        \hline
        \begin{tabular}{@{}c@{}}
           $\delta'$    \\
             (wall frame) 
        \end{tabular}
         & \begin{tabular}{@{}c@{}}Distance at which the plasma has \\[-1.7mm] reached its asymptotic equilibrium state\end{tabular}  & \begin{tabular}{@{}c@{}} $\delta'\gg \max(\gamma_w L_{\rm MFP},L_w)$ \\[-1.5mm] and $\delta'\ll \gamma_w R_{\rm bubble}$\end{tabular}  \\
        \hline
    \end{tabular}
    \caption{Summary of the different length scales relevant for the phase transition dynamics. $m_\phi$ is the scalar field's mass and $g^n$ is some power of the coupling constant. We have also indicated in which frame these quantities are defined.  } 
    \label{tab:scales}
\end{table}

The remainder of this article is as follows. In the next section, we give a brief review of the general aspects of bubble wall dynamics and hydrodynamics. In Sec.~\ref{sec:Bounds} we discuss the two approximations used in this paper. We derive the ballistic distribution functions with hydrodynamics correctly integrated and derive a new effective matching condition for the ballistic approximation. In Sec.~\ref{sec:proof}, we argue that the LTE and ballistic limits establish bounds on the wall friction and velocity. In Sec.~\ref{sec:numerics}, we present our numerical results both for a model-independent analysis and for an example model. We conclude in Sec.~\ref{sec:Conc}. For completeness, some technical details are left in appendices. Throughout we use the metric signature $(+,-,-,-)$.

\section{A brief review of bubble wall dynamics and hydrodynamics}
\label{sec:review}

\begin{figure}[ht]
    \centering
    \includegraphics[width=0.7\linewidth]{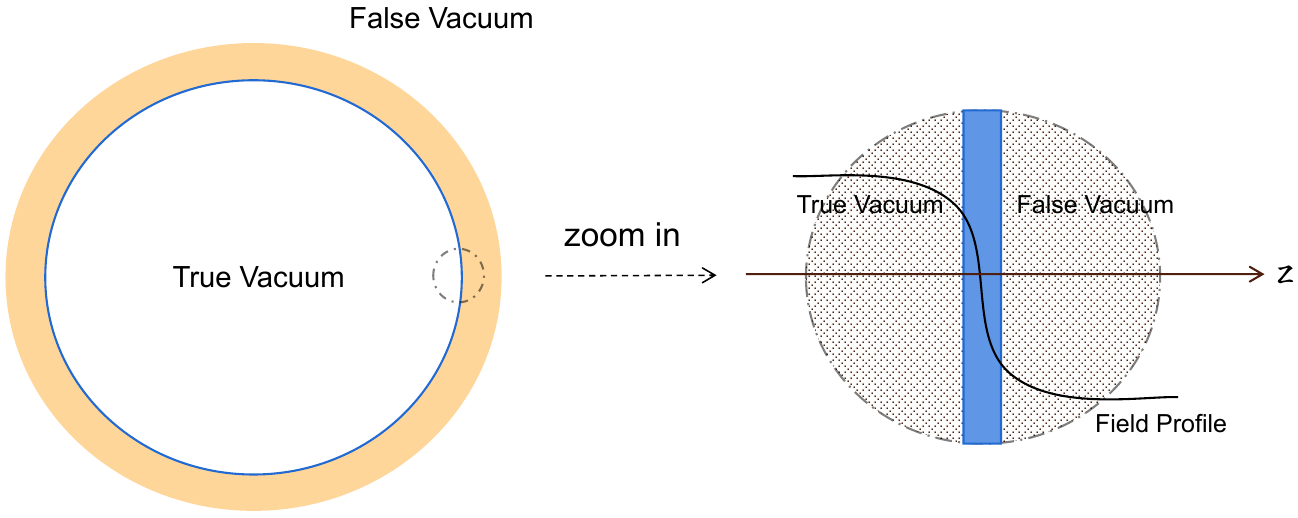}
    \caption{Sketch of bubble wall dynamics using the deflagration mode as an example. The bubble wall and shock front are viewed as having zero size at the hydrodynamic scale. The hydrodynamic quantities $T$ and $v$ are discontinuous at these fronts. To study the matching conditions, one then needs to zoom in on these fronts.
    }
    \label{fig:bubble_wall_dynamics}
\end{figure}

We illustrate the bubble wall dynamics in Fig.~\ref{fig:bubble_wall_dynamics}, taking a deflagration as an example. The relevant quantities for a hydrodynamic description are the fluid temperature and velocity distributions, $T(\xi)$ and $v(\xi)$, where $\xi=r/t$ with $r$ the radial coordinate and $t$ the time since nucleation. 
Since there is no other scale in the problem, the solutions for $v$ and $T$ are self-similar, i.e., they only depend on $\xi$. The bubble wall and the shock front can be viewed as having zero size at the hydrodynamics scale. $T(\xi)$ and $v(\xi)$ are continuous, except at the bubble wall and the shock front. The quantities on both sides of these fronts (either the phase front, i.e. bubble wall, or shock front) are therefore related by matching conditions. To study the matching conditions at the wall and to determine the wall velocity, one needs to zoom in on the fronts to a scale where the wall has a finite width. One uses the coupled EoMs between the scalar field and the plasma to find the matching conditions. There are two standard (hydrodynamic) matching conditions given by the conservation of the total energy-momentum tensor. For both of the two approximations adopted in this work, there is another matching condition with which one can fully determine the system for a given nucleation temperature $T_n$.

\subsection{Coupled scalar and plasma equations of motion}

The analysis of the friction on bubble walls is usually based on the following coupled EoMs for the background field and plasma~\cite{Liu:1992tn, Moore:1995ua,Moore:1995si},
\begin{subequations}
\label{eq:eoms}
\begin{align}
&\Box\phi+\frac{\d V(\phi)}{\d\phi}+\sum_i\frac{\d m^2_i(\phi)}{\d\phi}\int \frac{\d^3{\bf p}}{(2\pi)^32E_i}\,f_i(p,x)=0\,,\label{eq:EoM-bubble}\\
&\frac{\d f_i}{\d t}=-\C[f]\,, \label{eq:Boltz-Eq}
\end{align}
\end{subequations}
where $\Box=\partial_\mu \partial^\mu$,  $f_i(p,x)$ are the particle distribution functions, and $E_i=\sqrt{\vec{p}^2_i+m_i^2}$ the particle energies. $V(\phi)$ is the zero-temperature potential which may include quantum corrections to the classical potential. Considering a bubble wall expanding in the $z$-direction, and working in the rest frame of the wall where the wall position is taken as $z=0$, one can obtain the driving and frictional pressures (see, e.g., Ref.~\cite{Ai:2024shx})
\begin{subequations}
    \begin{align}
     &\P_{\rm driving}=V(\phi_s)-V(\phi_b)\equiv \Delta V\,,
     \label{eq:Pdriving}\\
    &\P_{\rm friction}= - \int_{-\delta}^{\delta} \d z\, (\partial_z\phi)\left(\sum_i\frac{\d m^2_i(\phi)}{\d\phi}\int \frac{\d^3{\bf p}}{(2\pi)^32E_i}\,f_i(p,z)\right)\,,\label{eq:Pfriction}
    \end{align}
\end{subequations}
where $\phi_{s,b}$ are the field values at the symmetric and symmetry-broken minima, respectively.\footnote{Although generally, FOPT does not necessarily involve symmetry breaking, we will use the terminology of a symmetry-breaking FOPT throughout.} Above, $\delta \gg L_w$ is a length scale. Its value should be chosen such that $|\partial_z\phi/(\Delta\phi)^2|_{\pm \delta}\ll 1$, where $\Delta\phi=\phi_b-\phi_s$. For a wall described by a hyperbolic tangent, one may take, e.g., $\delta= \mathcal O(5 L_w) $. It can be effectively understood to be infinity in the context of the scalar field EoM.

One can actually use conservation of total energy-momentum to write the frictional pressure in a different way. The energy-momentum tensors for the scalar field and plasma read 
\begin{subequations}
\begin{align}
T_{\phi}^{\mu\nu}&=(\partial^\mu\phi)\partial^\nu\phi-g^{\mu\nu}\left(\frac{1}{2}(\partial\phi)^2-V(\phi)\right)\,,\\
T_f^{\mu\nu}&=(e_f+p_f)u^\mu u^\nu- g^{\mu\nu}p_f\,,
\end{align}
\end{subequations}
where $u^\mu$ is the fluid four-velocity, and $p_f,e_f$ are the fluid contribution to the pressure and energy density. Note that the perfect fluid form of the plasma energy-momentum tensor already assumes LTE. Since these energy-momentum tensors will be used to derive the two standard matching conditions, we will see how those can be properly interpreted even though the plasma is out-of-equilibrium across the wall.
Then 
\begin{align}
    \partial_\mu T_\phi^{\mu\nu}=(\partial^\nu \phi) \left[\Box \phi+V'(\phi)\right]=-(\partial^\nu \phi)\sum_i\frac{\d m^2_i(\phi)}{\d\phi}\int \frac{\d^3{\bf p}}{(2\pi)^32E_i}\,f_i(p,x)\,,
\end{align}
where in the last equality we have used Eq.~\eqref{eq:EoM-bubble}.
Taking $\nu=z$, using total energy-momentum conservation, and working in the rest frame of the wall, one has
\begin{align}
    \partial_z T_f^{zz}= -(\partial_z \phi)\sum_i\frac{\d m^2_i(\phi)}{\d\phi}\int \frac{\d^3{\bf p}}{(2\pi)^32E_i}\,f_i(p,z)\,.
\end{align}
Therefore, 
\begin{align}
\label{eq:Pfriction-Tfzz}
    \P_{\rm friction}=\int_{-\delta}^{\delta} \d z\, \partial_z T_f^{zz}=T_f^{zz}|_{\rm in\ front\ of\ the\ wall}- T_f^{zz}|_{\rm behind\ the\ wall}\equiv\Delta T_f^{zz}\,.
\end{align}
On the other hand, we know
\begin{align}
\label{eq:Tfzz}
    T_f^{zz}=\sum_i\int\frac{\d^3 \vecp}{(2\pi)^3}\frac{(p^z)^2}{E_i} f_i(p,z)\,.
\end{align}
Once we know $f_i$ obtained from solving the Boltzmann equation, we can substitute them into Eqs.~\eqref{eq:Tfzz},~\eqref{eq:Pfriction-Tfzz} to obtain $\P_{\rm friction}$ without explicit dependence on $\phi$.
Note that, in calculating $T_f^{zz}$, one should only include the degrees of freedom that respond to the passage of the wall quickly enough so that they are not completely “invisible” to the wall. A condition on these degrees of freedom could be that their mean free path should be much smaller than the bubble radius (or the length of the shock wave or rarefaction wave), $L_{\rm MFP}\ll R_{\rm bubble}$.

If the interactions among the particles are very efficient, the plasma can remain close to thermal equilibrium during the bubble expansion.\footnote{Whether the plasma can remain close to LTE also depends on the strength of the force imposed from the wall on the particles.} 
In such a case, $f_i=f_i^{\rm eq}$ and one does not need to solve the Boltzmann equation. This is the LTE approximation. This gives the lower bound of the friction on the wall and hence the upper bound of its velocity. On the other hand, if the interactions among the particles are extremely weak, one can take the collision term to be zero in Eq.~\eqref{eq:Boltz-Eq} and particles in the plasma do not collide with each other while passing through the wall. Again, the Boltzmann equations become very simple and solutions can be obtained straightforwardly. This is the ballistic approximation. These two approximations dramatically simplify the problem but meanwhile, provide the lower and upper bounds on the friction and hence are still very important.

Mathematically, the conditions for the LTE and ballistic approximation to be valid are
\begin{subequations}
\begin{align}
&L_w/\gamma_w \gg L_{\rm MFP} \propto \frac{1}{\Gamma} \,,\qquad \text{(LTE condition)}\\
\label{eq:ballistic_condition}
 &L_w/\gamma_w \ll L_{\rm MFP} \propto \frac{1}{\Gamma} \,,\qquad \text{(ballistic condition)}
\end{align}
\end{subequations}
where $L_w$ is the wall width {\it in the wall rest frame}, $L_{\rm MFP}$ is the mean free path {\it in the plasma frame}, $\Gamma$ characterizes the collision rate between the particles, and $\gamma_w\equiv 1/\sqrt{1-v_w^2}$ is the Lorentz boost factor corresponding to $v_w$. Furthermore, if $L_{\rm MFP}\ll R_{\rm bubble}$, which is usually satisfied,\footnote{For example, for an electroweak PT with $T_n \sim 100 \, {\rm GeV}$, the mean free path is set by the weak interaction, while $1/(R_{\rm bubble} H) \sim \beta/H \sim 100$ and the number of relativistic degrees of freedom $g_* \sim 100$. The scales $L_{\rm MFP}$ and the bubble radius at the moment of collision are separated by about 14 orders of magnitude.} the particles perturbed by the wall's passage can relax to equilibrium sufficiently fast inside and outside the bubble. Therefore, one can assume a well-defined temperature in front of and behind the wall. We summarise the approximations used in this paper in Table~\ref{tab:approximations}.

\begin{table}[ht]
    \centering
    \begin{tabular}{|c||c|c|}
        \hline
        Approximation & Scale hierarchy & Reference \\
         \hline\hline
        LTE & $R_{\rm bubble}\gg L_w/\gamma_w\gg L_{\rm MFP}$ & \cite{Balaji:2020yrx,Ai:2021kak,Ai:2023see} \\
         \hline
        Ballistic (this work) & $R_{\rm bubble}\gg L_{\rm MFP}\gg L_w/\gamma_w$ & Section \ref{sec:ballistic} \\
         \hline
        Ballistic (previous literature) & $L_{\rm MFP}\gg R_{\rm bubble}\gg L_w/\gamma_w$ & \cite{Liu:1992tn,BarrosoMancha:2020fay} \\
         \hline
    \end{tabular}
    \caption{Summary of the different approximations found in the literature and studied in this work, and the scale hierarchies at which they are valid. }
    \label{tab:approximations}
\end{table}

There are three ways to apply the ballistic approximation. The first way is commonly considered in the literature~\cite{Bodeker:2009qy,Bodeker:2017cim,Hoche:2020ysm,Azatov:2020ufh,Gouttenoire:2021kjv,GarciaGarcia:2022yqb,Ai:2023suz}; one fixes the collision strength and hence $L_{\rm MFP}$ to be the one determined by the model under study, and  condition~\eqref{eq:ballistic_condition} is satisfied for sufficiently large $\gamma_w$. Therefore, the ballistic approximation is used for ultrarelativistic bubble walls, i.e. in the large $\gamma_w$ limit. This is a faithful application of this approximation, but it only tells us the friction in the ultrarelativistic regime. 
In the other two approaches, one takes the limit $\Gamma\rightarrow 0$  for given $\gamma_w$. In this way, one does not faithfully derive the wall velocity for the considered model but uses the ballistic approximation to derive the upper bound of the friction and thus the lower bound of the wall velocity (the same holds for the use the LTE approximation in this work). 
In the approach of~\cite{Liu:1992tn,BarrosoMancha:2020fay}, the plasma temperature and velocity were taken as constant. In this work, on the other hand we also consider inhomogeneous plasma temperature and velocity distributions, thus accounting for hydrodynamic obstruction~\cite{Konstandin:2010dm,Cline:2021iff,Laurent:2022jrs,Ai:2023see,Ai:2024shx}. We shall observe a higher pressure barrier in the ballistic approximation than in the LTE approximation. This is a crucial observation made in this work.

\subsection{The standard matching conditions and fluid equations}

\paragraph{Two standard matching conditions} 

To study the matching conditions for the hydrodynamic quantities across the wall, it is convenient to work in the rest frame of the bubble wall (see the right panel of Fig.~\ref{fig:bubble_wall_dynamics}). 
The matching conditions are usually formulated in terms of the energy density and pressure with the tree-level scalar potential energy absorbed: $e=e_f+V(\phi)$, $p=p_f-V(\phi) \equiv -V_{\rm eff}(\phi,T)$.  Note that the fluid enthalpy is $\omega=e_f+p_f=e+p$. From energy-momentum conservation, we have in equilibrium
\begin{subequations}
\begin{align}
    &\omega(z)\gamma^2(z) v(z)={\rm const}\, ,\\
    \label{eq:em-con-2}
    &\omega(z)\gamma^2(z) v^2(z)+\frac{1}{2}(\partial_z\phi(z))^2+p(z)={\rm const}\, ,
\end{align}
\end{subequations}
where $v >0$ is defined through $u^\mu=\gamma(1,0,0,-v)$ and $\gamma(v)=1/\sqrt{1-v^2}$. To get the matching conditions, we take $z=\pm \delta'$, (c.f. Table~\ref{tab:scales}). Since the above equations are valid only for LTE (due to the use of the perfect fluid energy-momentum tensor), $\delta'$ has to be chosen to be much larger than the mean free path in the wall frame $\gamma_w L_{\rm MFP}$ (which should really be understood as the mean free path of the particle with the smallest interaction rate). This way, all particles relax to equilibrium at $\pm \delta'$. On the other hand, $\delta'$ should be much smaller than the macroscopic scale so that it can be essentially viewed as infinitesimally thin at the hydrodynamics scale. Therefore, we have
$\gamma_w R_{\rm bubble} \gg\delta' \gg \gamma_w L_{\rm MFP}$. 
Then we obtain
the following two well-known matching conditions
\begin{subequations}
\label{eq:junctionAB}
\begin{align}
    &\omega_+\gamma_+^2v_+=\omega_-\gamma_-^2v_-\, ,\label{eq:conditionA}\\
    &\omega_+\gamma_+^2v_+^2+p_+=\omega_-\gamma_-^2v_-^2+p_-\, ,\label{eq:conditionB}
\end{align}
\end{subequations}
where a subscript ``$\pm$'' is used to denote quantities in front of/behind the bubble wall. To be clear, $e_+=e_s(T_+)$, $e_-=e_b(T_-)$ and similarly for $p_{\pm }$. Now, the thermodynamic quantities are understood to be the thermal equilibrium ones in the above matching conditions.

Eqs.~\eqref{eq:junctionAB} are definite only when $p$ is a known function of $T$. This, in principle, requires knowledge about the effective potential of a given model $p=-V_{\rm eff}(\phi, T)$ where we have included the $\phi$-independent term in $V_{\rm eff}$. We are only interested in the pressures in the symmetric phase $\phi_s(T)$ and the broken phase $\phi_b(T)$, which are determined by minimizing the potential for a given $T$. Therefore, $p$ can be understood as a function of the temperature only. The other thermodynamic quantities of interest can then be obtained from $p(T)$, e.g.,
\begin{equation}
\label{eq:thermo-quantities-from-p}
    e= T\frac{\d p}{\d T} - p\,, \qquad 
    \omega = T\frac{\d p}{\d T}\,, \qquad s = \frac{\omega}{T}\,,\qquad  c^2= \frac{\d p/\d T}{\d e/\d T}\,, 
\end{equation}
where $s$ and $c$ are the entropy density and sound speed respectively. With a known function $p(T)$, Eqs.~\eqref{eq:junctionAB} are two constraint equations for five quantities $v_w$, $T_\pm$ and $v_\pm$. Since $T_+$, $v_+$ can be related to the nucleation temperature $T_n$ and wall velocity $v_w$, one actually has only one unknown among the five quantities $\{v_w, T_\pm, v_\pm\}$ after imposing Eqs.~\eqref{eq:junctionAB}. In Section~\ref{sec:Bounds} we demonstrate how to supplement the system with a third matching condition, such that all quantities are fixed.

To perform a model-agnostic analysis, one usually replaces an exact $p(T)$ with an assumed equation of state (EoS). Often-used examples are the bag EoS or its generalisation, the template model~\cite{Leitao:2014pda}. We briefly review the model-independent analysis of the hydrodynamic matching relations~\cite{Ai:2023see} based on the template model in Appendix~\ref{app:template}, which we also use to provide a proof that the LTE approximation provides a lower bound on the friction in Section~\ref{sec:LTE_as_bound_template}.

\paragraph{Fluid equations}

Away from the bubble wall $|z| \sim R_{\rm bubble}$ $> \delta, \delta'$), the fluid properties are determined by the fluid equations: $u_\nu \partial_\mu T^{\mu\nu}_f=0, \Bar{u}_\nu\partial_\mu T^{\mu\nu}=0\,,$
where $\Bar{u}^\mu$ is the normalized vector orthogonal to $u^\mu$. For these equations, it is convenient now to work in the rest frame of the bubble center. Then we have $u^\mu=\gamma(1,\vec{v})$ and $\Bar{u}^\mu=\gamma(v,\vec{v}/v)$. In spherical coordinates, $u^\mu=(\gamma,\gamma v,0,0)$ and $\Bar{u}^\mu=(\gamma v,\gamma,0,0)$. Since there is no characteristic scale in the problem, the solution should depend only on the dimensionless variable $\xi=r/t$. The fluid equations can be written in a more explicit form, see e.g. Ref.~\cite{Espinosa:2010hh}.

The fluid dynamics scale is characterised by the bubble radius $R_{\rm bubble}$ while the bubble wall dynamics scale is given by the wall width $L_w$. The condition $L_w \ll \delta\ll \gamma_w R_{\rm bubble}$ for Eqs.~\eqref{eq:Pfriction} and~\eqref{eq:Pfriction-Tfzz} means that although $\delta$ can be effectively taken to be infinity in these equations, it can also be effectively viewed as $0^+$ in terms of the dimensionless variable $\xi$ in the plasma frame. Therefore, in $\xi$-space, one essentially integrates from $\xi_w -0^+$ to $\xi_w +0^+$ to obtain the friction. This is reasonable as the friction on the wall only depends on the local state of the plasma near the wall.\footnote{In contrast, Refs.~\cite{Wang:2022txy,Wang:2023kux} define the friction as an integral from $\xi=0$ to $\xi=1$ so that the wall friction receives additional contributions from regions far away from the wall.}

\section{Bubble wall velocity in the LTE and ballistic approximations}
\label{sec:Bounds}

We have just discussed the two well-known hydrodynamic matching relations. Now, we will discuss how we can obtain a third matching relation in the limit of LTE or the ballistic approximation, which allows us to determine $v_w$.

\subsection{The LTE approximation}

\label{sec:LTE}

In the LTE approximation, there is an additional matching condition due to the entropy conservation across the wall~\cite{Ai:2021kak}
\begin{equation} 
\label{eq:LTE-matching_1}
   \partial_\mu (s u^\mu) = 0\quad\Rightarrow\quad  s_+ \gamma_+ v_+ =s_- \gamma_- v_-\,,
\end{equation}
where $s\equiv \partial p/\partial T$ is the entropy density.
Using Eq.~\eqref{eq:conditionA}, it can be also written as
\begin{align}
\label{eq:LTE-matching-condition}
    \gamma_+T_+=\gamma_- T_-  \qquad \text{(LTE\ matching\ condition)} \,.
\end{align}
This provides the third matching condition as an addition to the previous two given in Eqs.~\eqref{eq:conditionA} and~\eqref{eq:conditionB}. And one can fully determine the wall velocity for any specified effective potential $V_{\rm eff}$ and a given nucleation temperature $T_n$. 
In Section \ref{sec:LTE_as_bound_template} we present a proof that the LTE limit gives the upper bound on the wall velocity using the template model EoS.

\subsection{The ballistic approximation}
\label{sec:ballistic}

In the wall frame, a LTE distribution function takes the form 
\begin{align}
\label{eq:LTE-distribution}
    f^{\rm eq}(p^z,z;\vecp_\perp)=\frac{1}{\e^{p_\mu u^\mu/T}\pm 1}=\frac{1}{\e^{\beta\gamma (E+v p^z)}\pm 1}\,,
\end{align}
where $\beta\equiv 1/T$, $E=\sqrt{\vecp^2+m^2(z)}$. Note that all the quantities except for $\vecp$ should be understood as a function of $z$. 

In the ballistic approximation, particles {\it across} the wall are not in LTE. However, one expects that the fluid in front of the wall and behind the wall can have a well-defined temperature for particles moving towards the wall,  given by $T_+$ and $T_-$, respectively. Furthermore, the bulk velocities of the fluid in front of and behind the wall are given by $v_+, v_-$ respectively. 
The incident modes (modes that move towards the wall) are in thermal equilibrium at $z \rightarrow \pm \delta$ (in the rest of this section, we will simply replace $\delta$ with $\infty$ to follow the convention in the literature), 
i.e., 
\begin{subequations}
\label{eq:bdy}
\begin{align}
\label{eq:bdy1}
     f_\infty^{\leftarrow} &=\frac{1}{\e^{\beta_+\gamma_+(E+v_+ p^z)} \pm 1}\,, \quad (p^z<0, z\rightarrow \infty, m=m_+)\,,\\
     \label{eq:bdy2}
     f^{\rightarrow}_{-\infty} &= \frac{1}{\e^{\beta_- \gamma_- (E+v_-p^z)}\pm 1}\,,\quad (p^z>0, z\rightarrow -\infty, m=m_->m_+)\,,
\end{align}
\end{subequations}
where we have denoted $m_\pm=m(z\rightarrow \pm \infty)$. Above and in the following, we assume that the particles gain mass when they enter the bubble. In Appendix~\ref{app:mass-loss}, we extend the analysis to the case when particles lose mass as they enter the bubble.

The full distribution should be determined by the Liouville equation
\begin{align}
\label{eq:Liouville}
    \left(\frac{p^z}{E}\partial_z -\frac{\partial_z m^2 }{2E}\partial_{p^z}\right) f(p^z,z;\vecp_\perp)=0\,.
\end{align}
Using $\partial_z =(\partial_z m^2)\partial_{m^2}$, one has 
\begin{align}
    \frac{p^z(\partial_z m^2)}{E}\left(\partial_{m^2}-\partial_{p^{z2}}\right) f(p^z,z;\vecp_\perp)=0\,.
\end{align}
This simply means that $f$ is a function of the combination $m^2+p^{z2}$. As a consequence, it should be straightforward to obtain the solution for a given boundary condition.

The solution to the Liouville equation with the boundary conditions~\eqref{eq:bdy} can be interpreted physically~\cite{Liu:1992tn, BarrosoMancha:2020fay}. 
\begin{itemize}
    \item[(1)] Transmission from the symmetric phase ($t_+$):
    \begin{align}
        f^{t_+}(p^z,z;\vecp_\perp)=\frac{1}{\e^{\beta_+\gamma_+\left(E-v_+\sqrt{p^{z2}+m^2(z)-m_+^2}\right)}\pm 1}\,, \quad \left(p^z<-\sqrt{m_-^2-m^2(z)}\right)\,;
    \end{align}
    \item[(2)] Reflection ($r$): 
    \begin{align}
        f^r(p^z,z;\vecp_\perp)=&\frac{1}{\e^{\beta_+\gamma_+ \left(E-v_+\sqrt{p^{z2}+m^2(z)-m_+^2}\right)}\pm 1}\,,\notag\\&\qquad\qquad\qquad\qquad\qquad\left(-\sqrt{m_-^2-m^2(z)}<p^z<\sqrt{m_-^2-m^2(z)}\right)\,;
    \end{align}
    Note that above we do not have $\sqrt{p^{z2}+m^2(z)-m_+^2}$ multiplied with a factor of ${\rm sign}[p^z]$, in contrast to Ref.~\cite{BarrosoMancha:2020fay}.\footnote{This observation has also been made in Ref.~\cite{Wang:2024wcs}. However, our solutions differ from those given in Ref.~\cite{Wang:2024wcs}. We have taken into account the inhomogeneity in the fluid velocity so that $v_+$, $v_-$ enter the solutions. There seem to be also some sign differences between our solutions and those in Ref.~\cite{Wang:2024wcs}.}
    \item[(3)] Transmission from inside the bubble $(t_-)$:
    \begin{align}
        f^{t_-}(p^z,z;\vecp_\perp)=\frac{1}{\e^{\beta_-\gamma_-\left(E+v_-\sqrt{p^{z2}+m^2(z)-m_-^2}\right)}\pm 1}\,, \quad \left(p^z>\sqrt{m_-^2-m^2(z)}\right)\,.
    \end{align}
\end{itemize}
The above solution completes the distribution function for all the $p^z$-modes in front of the wall ($z\rightarrow\infty$) and behind the wall ($z\rightarrow-\infty$). Explicitly, at $z\rightarrow \infty$, we have 
\begin{equation}
\label{eq:f+inf}
f(p^z,\infty;\vecp_\perp) = \begin{cases}
\frac{1}{\e^{\beta_+\gamma_+\left(E+v_+ p^z\right)}\pm 1} \,,\qquad\qquad\ p^z<-{\sqrt{\Delta m^2}}\,,\quad &(\text{$t^+$-modes}) \\
\frac{1}{\e^{\beta_+\gamma_+\left(E- v_+ |p^z|\right)}\pm 1}\,,\quad -{\sqrt{\Delta m^2}}<p^z<{\sqrt{\Delta m^2}}\,,\quad &(\text{$r$-modes})\\
\frac{1}{\e^{\beta_-\gamma_-\left(E+v_-\sqrt{p^{z2}-\Delta m^2}\right)}\pm 1}\,,\ \ \, p^z> {\sqrt{\Delta m^2}}\,, \quad &(\text{$t^-$-modes})
\end{cases}\,,
\end{equation}
where $\Delta m^2 \equiv  m_-^2- m_+^2$. In general $\Delta m^2$ is not equal to $(\Delta m)^2\equiv (m_- -m_+)^2$ unless $m_+=0$.
At $z\rightarrow -\infty$, we have 
\begin{align}
\label{eq:f-inf}
    f(p^z,-\infty;\vecp_\perp)=
\begin{cases}
    \frac{1}{\e^{\beta_+ \gamma_+ \left(E-v_+\sqrt{p^{z2}+\Delta  m^2}\right) }\pm 1}\,,\quad p^z<0\,,\quad &(\text{$t^+$-modes})\\
    \frac{1}{\e^{\beta_-\gamma_- \left(E+v_-p^z\right)}\pm 1}\,,\quad\qquad\quad p^z>0\,,\quad &(\text{$t^-$-modes})
\end{cases}\,.
\end{align}
Note that behind the wall, there are no $r$-modes. This is reasonable as reflected particles do not enter the wall. We also note that $T_+,T_-$ are not defined for all the modes in front of/behind the wall. Let us stress that there is an important difference between the approach we take and the ballistic computation done in~\cite{Liu:1992tn, BarrosoMancha:2020fay}; we distinguish the fluid velocity and temperature in front of and behind the bubble wall, whereas in~\cite{Liu:1992tn, BarrosoMancha:2020fay}, these are taken to be constant at $v_w$ and $T_n$ respectively. As we will see explicitly in Sec.~\ref{sec:example-model}, this has a significant effect on the pressure, especially at velocities close to the Jouguet velocity (the transition from hybrid solutions to detonations). With the assumption of a constant fluid velocity and temperature, the ballistic approximation does \emph{not} provide an upper bound on the friction. On the other hand, the ballistic approximation studied here is the limit one would obtain when solving the Boltzmann equation and slowly turning off the collisions in the plasma ($\mathcal{C}[f]\to0$) which as we will show, makes it indeed the upper bound on the friction and thus lower bound on the wall velocity.

To compute the pressure on the wall, it is convenient to isolate the contribution from each mode. Using Eqs. (\ref{eq:Pfriction-Tfzz}) and (\ref{eq:Tfzz}), one can then write
\begin{align}
    \P^{\rm b}_{\rm friction} (T_+,T_-,v_+,v_-) = \P^{t_+} + \P^{t_-} + \P^r\,,
\end{align}
with
\begin{subequations}
\begin{align}
    \P^{t_+} &= \int\limits_{p^z<-{\sqrt{\Delta m^2}}}\frac{\d^3\vecp}{(2\pi)^3}\frac{(p^z)^2}{E_+}f^{t_+}(p^z,+\infty) - \int\limits_{p^z < 0}\frac{\d^3\vecp}{(2\pi)^3}\frac{(p^z)^2}{E_-}f^{t_+}(p^z,-\infty)\,,\\
    \P^{t_-} &= \int\limits_{p^z>{\sqrt{\Delta m^2}}}\frac{\d^3\vecp}{(2\pi)^3}\frac{(p^z)^2}{E_+}f^{t_-}(p^z,+\infty) - \int\limits_{p^z > 0}\frac{\d^3\vecp}{(2\pi)^3}\frac{(p^z)^2}{E_-}f^{t_-}(p^z,-\infty)\,,\\
    \P^r &= \int\limits_{|p^z|<{\sqrt{\Delta m^2}}}\frac{\d^3\vecp}{(2\pi)^3}\frac{(p^z)^2}{E_+} f^r(p^z,+\infty)\,,
\end{align}
\end{subequations}
where $E_{\pm}=E(z=\pm\infty)$ and we did not make the $\vecp_\perp$-dependence in the distribution functions explicit. These equations can be simplified by observing that
\begin{subequations}
\begin{align}
    &f^{t_+}(p^z,+\infty) = f^{\rm eq}(p^z,+\infty)\,, \quad &&f^{t_-}(p^z,-\infty) = f^{\rm eq}(p^z,-\infty)\,,\\
    &f^{t_+}(p^z,-\infty)=f^{\rm eq}(-\sqrt{p^{z2}+ \Delta m^2},+\infty)\,,\quad &&f^{t_-}(p^z,+\infty)=f^{\rm eq}(+\sqrt{p^{z2}- \Delta m^2},-\infty)
\end{align}
\end{subequations}
and $f^r(p^z,+\infty)=f^{\rm eq}({-}|p^z|,+\infty)$ to replace all the ballistic distribution functions with the corresponding equilibrium one. Then, with appropriate changes of variables, the pressures can be written as 
\begin{subequations}
\label{eq:P-ballistic-momentum-ex}
\begin{align}
    \P^{t_+} &= \int\limits_{p^z<-{\sqrt{\Delta m^2}}}\frac{\d^3\vecp}{(2\pi)^3}\frac{p^z}{E_+}\left(p^z +\sqrt{(p^z)^2-\Delta m^2}\right)f^{\rm eq}(p^z,+\infty)\,,\\
    \P^{t_-} &= \int\limits_{p^z>0}\frac{\d^3\vecp}{(2\pi)^3}\frac{p^z}{E_-}\left(\sqrt{(p^z)^2+\Delta m^2}-p^z\right)f^{\rm eq}(p^z,-\infty)\,,\\
    \label{eq:Pr-exact2}
    \P^r &= 2\int\limits_{-{\sqrt{\Delta m^2}} < p^z < 0}\frac{\d^3\vecp}{(2\pi)^3}\frac{(p^z)^2}{E_+}f^{\rm eq}(p^z, +\infty)\,.
\end{align}
\end{subequations}
The total frictional pressure in the ballistic approximation finally takes the simple form
\begin{align}
    \P^{\rm b}_{\rm friction} = \int \frac{\d^3\vecp}{(2\pi)^3}\frac{p^z}{E}\Delta p^z f^{\rm eq}\,,
\end{align}
where $\Delta p^z$ is the exchanged momentum with the wall and $f^{\rm eq}$ and $E$ are evaluated at $z\to +\infty$ for $r$ and $t^+$ modes and $z\to-\infty$ for the $t^-$ mode. This last equation has a clear physical interpretation: particles coming from $\pm\infty$ hit the wall with a flux $\frac{p^z}{E_\pm}f^{\rm eq}(p^z,\pm \infty)$ with each collision transferring the momentum $\Delta p^z$ to the wall. The final friction felt by the wall is then obtained by integrating the product of these two factors.

The integrals simplify if we approximate the Bose-Einstein/Fermi-Dirac distribution by the Boltzmann distribution. Integrating over $\vecp_\perp$ analytically and with appropriate changes of variables, we obtain
\begin{subequations}
\label{eq:P-ballistic-momentum-final-ex}
\begin{align}
    \P^{t_+} &= \frac{T_+^4}{4\pi^2 \gamma_+}\int\limits_{x>{\sqrt{\Delta m^2}}/T_+}\d x\, x\left(x-\sqrt{x^2-\Delta m^2/T_+^2}\right)\e^{-\gamma_+\left(\sqrt{x^2+m_+^2/T_+^2}-v_+x\right)}\,,\\
    \P^{t_-} &= \frac{T_-^4}{4\pi^2 \gamma_-}\int\limits_{x>0}\d x\, x\left(\sqrt{x^2+\Delta m^2/T_-^2}-x\right)\e^{-\gamma_-\left(\sqrt{x^2+m_-^2/T_-^2}+v_-x\right)}\,,\\
    \label{eq:Pr-approx}
    \P^r &= \frac{T_+^4}{2\pi^2\gamma_+}\int\limits_{0<x<\sqrt{\Delta m^2}/T_+}\d x\, x^2\,\e^{-\gamma_+\left(\sqrt{x^2+m_+^2/T_+^2}-v_+x\right)} \,.
\end{align}
\end{subequations}
We present the corresponding expressions of the various pressures for Bose-Einstein and Fermi-Dirac distributions in Appendix~\ref{app:pressures-formulae}.

At last, 
\begin{align}
\label{eq:ballistic_matching_condition}
    \Delta V= \P^{\rm b}_{\rm friction}(T_+,T_-,v_+,v_-) \qquad ({\rm ballistic\ matching\ condition})
\end{align}
effectively provides an additional matching condition for $\{T_+,T_-,v_+,v_-\}$ and hence we can solve the wall velocity as we do for the LTE case. In reality, there are typically multiple particle species that couple with the order-parameter scalar $\phi$. In that case, one needs to sum over all the contributions with parameters $m_{i,\pm}$. 

\paragraph{The ultrarelativistic limit} As mentioned earlier, the ballistic condition~\eqref{eq:ballistic_condition} itself does not necessarily require ultrarelativistic walls. However, it is interesting to see how the standard B\"odeker-Moore $1$-to-$1$ thermal friction, derived in the $\gamma_w\rightarrow\infty$ limit~\cite{Bodeker:2009qy}, can be recovered from the above formulae. In the limit $\gamma_w\rightarrow\infty$, the equilibrium distribution function~\eqref{eq:LTE-distribution} is a very narrow distribution peaked at $p^z=-\gamma T$ (see e.g.~\cite{Ai:2023suz}). Therefore, from the integral ranges in Eqs.~\eqref{eq:P-ballistic-momentum-ex}, one sees that $\P^{t_-}$ and $\P^r$ are suppressed compared with $\P^{t_+}$. Furthermore, for $\gamma_w\rightarrow \infty$, one can well assume that the bubble motion is in the detonation mode 
and thus $f^{\rm eq}$ in front of the wall is estimated at the nucleation temperature $T_+=T_n$. Since only the narrow region near $-\gamma_w T_n$ makes the dominant contribution to the integral of $\P^{t_+}$, one can Taylor expand $\sqrt{p^{z2}-\Delta m^2}\approx p^z-\frac{\Delta m^2}{2p^z}$, assuming $|p^z|\gg \Delta m^2$. Finally one gets
\begin{align}
\label{eq:frictonBM}
\P_{\gamma_w\rightarrow\infty}\approx \P^{t_+}&\approx \Delta m^2 \int\limits_{p^z<-\Delta m}\frac{\d^3\vecp}{(2\pi)^3 2E_+}f^{\rm eq}(p^z,+\infty;T_n)\notag\\
&\approx \Delta m^2 \int\limits\frac{\d^3\vecp}{(2\pi)^3 2E_+}f^{\rm eq}(p^z,+\infty;T_n)\\
&\!\!\stackrel{m_+=0}{=}\left\lbrace \Delta m^2 T^2/48,\quad \mathrm{fermions}\atop \Delta m^2 T^2/24,\quad \mathrm{bosons}\ \ \right. 
\end{align}
where the last expression is the standard Bödeker-Moore $1$-to-$1$ thermal friction $\P_{\rm BM}$ which is valid when $m_+=0$.

\section{Demonstration that LTE and ballistic provide bounds on the wall velocity}
\label{sec:proof}

Having discussed how the LTE and ballistic approximations simplify the computation of the bubble wall velocity, here we demonstrate that they indeed provide bounds on the wall velocity and friction. We will first prove that the LTE gives an upper bound of the wall velocity using the template model (cf. Appendix~\ref{app:template}), following Ref.~\cite{Ai:2024shx}. Then, we give a more general argument based on the linearized Boltzmann equation.

\subsection{LTE as an upper bound on the wall velocity in the template model}

\label{sec:LTE_as_bound_template}

In the general case where LTE is not maintained in the plasma, one cannot expect entropy to be conserved across the wall. Therefore, Eq.~\eqref{eq:LTE-matching_1} must be replaced by the more general condition of the non-negativity of entropy production~\cite{Laine:1993ey}
\begin{align}
\label{eq:entropycreation}
    v_-\gamma_- s_- - v_+\gamma_+ s_+ = v_+\gamma_+\Delta s \geq 0\,.
\end{align}
Here, $\Delta s$ quantifies the variation of entropy which depends on the deviation from equilibrium across the wall, making it a highly model-dependent quantity. Nevertheless, as the total entropy cannot decrease, $\Delta s$ is always positive, and this fact can be used to prove that the LTE limit gives an upper bound of the wall velocity. We will restrict our proof to the template EoS. 

Using Eq.~\eqref{eq:entropycreation} instead of~\eqref{eq:LTE-matching_1}, the temperature behind the wall can now be expressed as
\begin{align}
    T_-=\frac{\gamma_+ T_+}{\gamma_-(1+\sigma)}\,,
\end{align}
where we have defined 
\begin{equation}
\label{eq:sigma}
    \sigma\equiv\Delta s/s_+\geq 0\,.
\end{equation}
Substituting the above into Eq.~\eqref{eq:match0} to remove the dependence on $T_-$, one finally obtains
\begin{align}\label{eq:matchout}
    3\nu\alpha_+ v_+ v_- = \left[1-3\alpha_+ - \left(\frac{\gamma_+}{\gamma_-}\right)^\nu \frac{\Psi_+}{(1+\sigma)^\nu}\right] \left(1-\frac{v_+ v_-}{c_b^2}\right)\,,
\end{align}
where the corresponding quantities are defined in Appendix~\ref{app:template}.
The above equation is the same as the LTE matching equation~\eqref{eq:match2} with the simple substitution
\begin{align}
    \Psi\to \Psi_{\rm eff}=\frac{\Psi}{(1+\sigma)^\nu}\leq \Psi\,,
\end{align}
where the last inequality comes from $\sigma\geq0$.

We conclude that, within the hydrodynamic description, an out-of-equilibrium plasma can be treated mathematically like an LTE plasma, with an effective $\Psi_{\rm eff}$ which gets reduced compared with the physical $\Psi$ by the entropy production. Note that this is just a mathematical correspondence. In the presence of a deviation from LTE, $\Psi_{\rm eff}$ loses the original meaning as the ratio of the enthalpies in the broken and
symmetric phases. In LTE, a smaller value of $\Psi$ leads to a smaller wall velocity (see Ref.~\cite{Ai:2023see}), as it means more particles can interact with the wall leading to a higher friction force. Using the correspondence above, this means that entropy production typically makes the wall velocity smaller compared with the limit of no entropy production. Hence the LTE approximation ($\sigma=0$) gives the upper bound of the wall velocity. 

Interestingly, in the limit $\Psi\to 0$ where all the degrees of freedom become very massive and cannot enter the bubble wall, one can see that LTE always gives an exact result as $\Psi_{\rm eff}=\Psi=0$. This curious fact was already pointed out in Ref.\ \cite{Sanchez-Garitaonandia:2023zqz} and happens because in this limit, there is no plasma on the broken side which implies $s_-=0$. With that constraint, it is easy to see that the only non-negative value Eq.~\eqref{eq:entropycreation} can have is 0. Therefore, entropy must be conserved and LTE becomes exact. 

\subsection{Demonstration that LTE and ballistic approximations bound $v_w$ in the linearized Boltzmann equation}

The general Boltzmann equation for the deviation from equilibrium $\delta f=f-f^{\rm eq}$ of a single particle can be written as
\begin{align}
\label{eq:Boltzmann}
    \left(p^z\partial_z-\frac{1}{2}\partial_z(m^2)\partial_{p^z}\right) \delta f+\lambda\mathcal{C}[\delta f](z,p^z) = -\left(p^z\partial_z-\frac{1}{2}\partial_z(m^2)\partial_{p^z}\right) f^{\rm eq} \equiv S(z,p^z)\,,
\end{align}
where the parameter $\lambda$ represents the strength of the interactions involved in the collisions.\footnote{Note that $\lambda$ does not correspond to a physical quantity and should be set to 1 in a calculation. We introduce it here to study the variation of the solution with respect to the collision strength and to take the different limits.} The ballistic limit can be obtained by taking $\lambda\to0$, while LTE is obtained in the limit $\lambda\to\infty$.
In general, $\mathcal{C}[\delta f](z,p^z)$ is a nonlinear functional of $\delta f$ (which is ultimately a function of $z$ and $p^z$), but for simplicity, we will linearize it with respect to $\delta f$. We will also assume $\partial_z(m^2)<0$, such that the particle becomes heavier when entering the bubble.

This equation can be simplified with the change of variable $\rho = \sqrt{(p^z)^2+m^2}$, which leads to the equation 
\begin{align}
    {\rm sign}(p^z)\sqrt{\rho^2-m^2}\frac{\d}{\d z}\delta f (z)+\lambda\mathcal{C}[\delta f]\left(z,{\rm sign}(p^z)\sqrt{\rho^2-m^2}\right)=S\left(z,{\rm sign}(p^z)\sqrt{\rho^2-m^2}\right)\,,
\end{align}
where $\delta f(z)=\delta f(z,{\rm sign}(p^z)\sqrt{\rho^2-m^2(z)})$.
We thus see that in terms of the $\rho$ variable, the modes with positive and negative momentum are governed by two distinct Boltzmann equations. This can be written in a simpler form as 
\begin{align}
    \frac{\d}{\d z}\delta f(z)
    -\lambda \mathcal{H}[\delta f](z,\rho,{\rm sign}(p^z))=\mathcal{S}(z,\rho,{\rm sign}(p^z))\,,
\end{align}
with 
\begin{align}
    \mathcal{H}[\cdot](z,\rho,{\rm sign}(p^z)) &= -\frac{{\rm sign}(p^z)}{\sqrt{\rho^2-m^2}}\mathcal{C}[\cdot]\left(z,{\rm sign}(p^z)\sqrt{\rho^2-m^2}\right), \notag \\
    \mathcal{S}(z,\rho,{\rm sign}(p^z)) &= \frac{{\rm sign}(p^z)}{\sqrt{\rho^2-m^2}}S\left(z,{\rm sign}(p^z)\sqrt{\rho^2-m^2}\right)= -\frac{\d}{\d z} f^{\rm eq}(z)\,.
\end{align}

To continue, it will be convenient to represent functions of $\rho$ and ${\rm sign}(p^z)$ as states of a vectorial space $\mathcal{V}$. The solution $\delta f(z,\rho,{\rm sign}(p^z))$ is now represented by a state $|{\delta f(z)}\rangle\in \mathcal{V}$, and similarly $\mathcal{S}(z,\rho,{\rm sign}(p^z))\to|{\mathcal{S}(z)}\rangle \in\mathcal{V}$. The parameter $z$ here is analogous to the time parameter in the Schr\"{o}dinger equation. Finally, the functional $\mathcal{H}[\cdot](z,\rho,{\rm sign}(p^z))$ is now represented by a map $\mathcal{H}(z)$ from $\mathcal{V}$ to $\mathcal{V}$.

The Boltzmann equations can then be expressed as 
\begin{align}
\label{eq:boltzmannVector}
    \frac{\d}{\d z}|{\delta f(z)}\rangle -\lambda\mathcal{H}(z)|{\delta f(z)}\rangle =|{\mathcal{S}(z)}\rangle \,.
\end{align}
To solve this equation, it will be useful to define the projection operators $\mathcal{P}_\pm$ that project out the negative and positive eigenvalues of $\mathcal{H}$. In other words, $\mathcal{H}_+\equiv \mathcal{P}_+\mathcal{H}=\mathcal{H}\mathcal{P}_+$ is positive definite and $\mathcal{H}_-\equiv \mathcal{P}_-\mathcal{H}=\mathcal{H}\mathcal{P}_-$ is negative definite. They also satisfy $\mathcal{P}_\pm^2=\mathcal{P}_\pm$, $\mathcal{P}_+\mathcal{P}_- = 0$, and since $\mathcal{H}$ has no zero eigenvalues,\footnote{The nonlinear collision operator \emph{does} have zero eigenvalues, as it must satisfy $\mathcal{C}\ket{f^{\rm eq}}=0$. 
After linearising this operator, this equation becomes $\mathcal{C}^{\rm linear}\ket{\delta f=0}=0$. However, a zero state cannot be an eigenstate. Furthermore, zero eigenvalues lead to divergent solutions (see Eqs.~\eqref{eq:boltzmannSolution}), hence cannot exist.} $\mathcal{P}_+ + \mathcal{P}_- = \mathds{1}$. Applying these operators on Eq.~(\ref{eq:boltzmannVector}), the positive and negative eigenmodes decouple as
\begin{align}
\label{eq:boltzmannProjected}
    \frac{\d}{\d z}|{\delta f_\pm(z)}\rangle -\lambda\mathcal{H}_\pm(z)|{\delta f_\pm(z)}\rangle =|{\mathcal{S}_\pm(z)}\rangle \,,
\end{align}
where $|{\delta f_\pm}\rangle =\mathcal{P}_\pm|{\delta f}\rangle $ and $|{\mathcal{S}_\pm}\rangle=\mathcal{P}_\pm|{\mathcal{S}}\rangle$. Physically, $|\delta f_\pm \rangle$ are deviations from equilibrium generated by particles coming from $\pm \infty$, respectively.

Let us now define the evolution operators $U_\pm(z,z')$ in terms of Dyson series as
\begin{align}
U_\pm(z,z')&=\mathcal{T}\exp\left(\lambda\int_{z'}^z\! \d y\, \mathcal{H}_\pm(y)\right)\nn \\
    &= \sum_{n=0}^\infty\frac{1}{n!}\mathcal{T}\left(\lambda\int_{z'}^z\! \d y\,\mathcal{H}_\pm(y)\right)^n\nn\\
    &= \mathds{1} + \lambda\int_{z'}^z\! \d y\,\mathcal{H}_\pm(y)+\lambda^2\int_{z'}^z\! \d y\int_{z'}^y\! \d y'\,\mathcal{H}_\pm(y)\mathcal{H}_\pm(y')+\cdots\,,
\end{align}
where $\mathcal{T}$ is the time-ordering operator (or position-ordering here), which arranges the $\mathcal{H}_\pm$ operators following it in order of decreasing distance from $z'$, such that the one most on the left is evaluated at the point farthest from $z'$, and the one on the right is evaluated at the point closest to $z'$. The evolution operators have the important properties that $U_\pm(z,z)=\mathds{1}$ and that their derivative with respect to $z$ is
\begin{align}
    \partial_z U_\pm(z,z') = \lambda\mathcal{H}_\pm(z)+\lambda^2\mathcal{H}_\pm(z)\int_{z'}^z\! \d y\,\mathcal{H}_\pm(y)+\cdots = \lambda\mathcal{H}_\pm(z)U_\pm(z,z')\,.
\end{align}

Using these properties, it becomes straightforward to show that the solutions of Eqs.~(\ref{eq:boltzmannProjected}) are
\begin{subequations}
\label{eq:boltzmannSolution}
\begin{align}
    \label{eq:boltzmannSolutionPlus}
    |{\delta f_+(z)}\rangle &= -\int_z^\infty\! \d z'\, U_+(z,z')|{\mathcal{S}_+(z')}\rangle \,, \\
    \label{eq:boltzmannSolutionMinus}
    |{\delta f_-(z)}\rangle  &= \int_{-\infty}^z\! \d z'\, U_-(z,z')|{\mathcal{S}_-(z')}\rangle \,.
\end{align}
\end{subequations}
Notice that the $U_\pm$ appearing in these solutions are both exponentials of negative definite operators, which guarantees that $U_\pm$ are positive definite and that the integrals converge (because $U_\pm(z,\mp\infty)=U_\pm(\pm\infty,z)=0$). Finally, the complete solution of the Boltzmann equation (\ref{eq:boltzmannVector}) is
\begin{align}
    |{\delta f(z)}\rangle = |{\delta f_+(z)}\rangle +|{\delta f_-(z)}\rangle = -\int_z^\infty\! \d z'\, U_+(z,z')|{\mathcal{S}_+(z')}\rangle +\int_{-\infty}^z\! \d z'\, U_-(z,z')|{\mathcal{S}_-(z')}\rangle \,.
\end{align}

We now wish to demonstrate how $|{\delta f(z)}\rangle$ varies with the interaction's strength, parametrized by $\lambda$. It is straightforward to show that the derivative of the evolution operators with respect to $\lambda$ is
\begin{align}
    \frac{\partial}{\partial\lambda}U_\pm(z,z') &= \int_{z'}^z\! \d y\, \mathcal{T}\left[\mathcal{H}_\pm(y)U_\pm(z,z')\right].
\end{align}
Note that, from the definition of the time-ordering operator $\mathcal{T}$, one has the relation $\mathcal{T}[\mathcal{H}_\pm(z),\mathcal{H}_\pm(z')] \allowbreak =0$. This implies the property $U_\pm(z,z'')=U_\pm(z,z')U_\pm(z',z'')$. Using this relation, the variation of the solutions with respect to $\lambda$ is
\begin{align}
    \frac{\partial}{\partial\lambda}|{\delta f_+(z)}\rangle  &= -\int_z^\infty\!\d z'\int_{z'}^z\! \d y\, \mathcal{T}\left[\mathcal{H}_+(y)U_+(z,z')\right] |{\mathcal{S}_+(z')}\rangle \nn\\
    &=-\int_z^\infty \!\d z' \left\lbrace -\partial_{z'}\left[ \int_{z'}^z\!\d y\int_{z'}^\infty\!\d y'\,\mathcal{T}[\mathcal{H}_+(y)U_+(z,y')]\ket{\mathcal{S}_+(y')} \right] \right. \nn\\
    &\hspace{25mm}\left. -\int_{z'}^\infty\!\d y'\,\mathcal{T}[\mathcal{H}_+(z')U_{+}(z,y')]\ket{\mathcal{S}_+(y')}\right\rbrace\nn\\
    &= \int_z^\infty\!\d z'\int_{z'}^\infty\! \d y'\, U_+(z,z')\mathcal{H}_+(z')U_+(z',y')|{\mathcal{S}_+(y')}\rangle \nn\\
    &= -\int_z^\infty\! \d z'\, U_+(z,z')\mathcal{H}_+(z')|{\delta f_+(z')}\rangle\,,
\end{align}
where the first term of the second line is zero because it is a total derivative vanishing at the boundaries, and we used Eq.~(\ref{eq:boltzmannSolutionPlus}) to get the fourth line. Similarly, we get
\begin{align}
\label{eq:lambdaVariationMinus}
    \frac{\partial}{\partial\lambda}|{\delta f_-(z)}\rangle  = \int_{-\infty}^z\! \d z'\, U_-(z,z')\mathcal{H}_-(z')|{\delta f_-(z')}\rangle \,.
\end{align}
These last two equations can be combined as
\begin{align}
\label{eq:lambdaVariation}
    \frac{\partial}{\partial\lambda}|{\delta f(z)}\rangle = -\int_{-\infty}^\infty\! \d z'\, |U(z,z')\mathcal{H}(z')|\, |{\delta f(z')}\rangle \,,
\end{align}
with $|U(z,z')\mathcal{H}(z')|=U_+(z,z')\mathcal{H}_+(z')\Theta(z'-z)-U_-(z,z')\mathcal{H}_-(z')\Theta(z-z')$, which is a positive definite operator.

Going back to the representation in terms of functions of $\rho$ and ${\rm sign}(p^z)$, $|U\mathcal{H}|$ becomes a functional which, due to its positive definiteness, maps every positive function to another positive function and vice versa. Therefore, Eq.~(\ref{eq:lambdaVariation}) implies that as we increase $\lambda$, the solution becomes closer and closer to zero, which corresponds to equilibrium.

\begin{figure}
    \centering
    \vspace{-1.3cm}
    \footnotesize{\hspace{0.1\linewidth}$m_+=0$, $m_-=3T_n$, $\alpha_n=0.3$, $b=1$ \hspace{0.15\linewidth} $m_+=0$, $m_-=T_n$, $\alpha_n=0.005$, $b=0.1$} \\
    \includegraphics[width=0.5\linewidth]{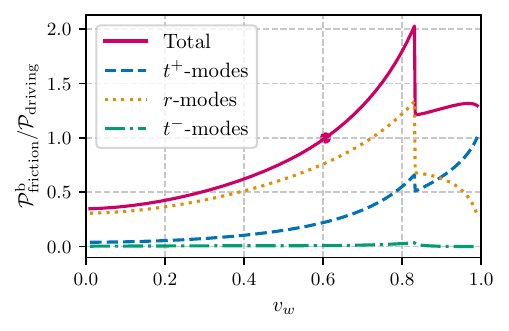}\includegraphics[width=0.5\linewidth]{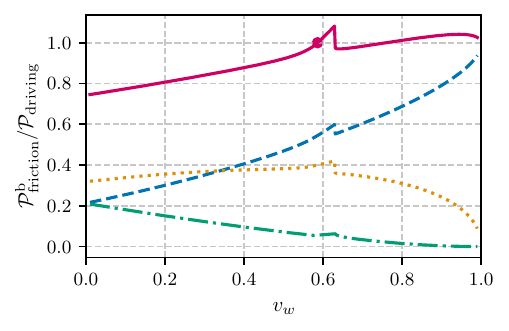}
    \caption{Different contributions to the ballistic pressure for a strong (left) and a weak (right) PT in the model presented in Section \ref{sec:model-independent-ballistic}. In both cases, the $t^-$-modes (which correspond to $\delta f_+$) give the smallest contribution. The points show the solutions satisfying $\mathcal{P}_{\rm friction}^{\rm b}/\mathcal{P}_{\rm driving}=1$.
    }
    \label{fig:ballisticPressures}
\end{figure}

However, in general, the solution of the Boltzmann equation $\delta f$ is not positive for all values of $z,p^z$. In fact, because $U_\pm$ are positive definite operators and because $\mathcal{S}=-\d f^{\rm eq}/\d z < 0$ (the particle is more massive in the bubble at $z\to -\infty$ and therefore Boltzmann suppressed), one can see from Eqs.~(\ref{eq:boltzmannSolution}) that $\delta f_+ > 0$ and $\delta f_-<0$. In practice, the pressure on the wall is dominated by the contribution from $\delta f_+$, as it corresponds to particles coming from the symmetric phase that are either transmitted to the broken phase or reflected, while $\delta f_-$ contains particles transmitted from the broken to the symmetric phases. The latter are typically Boltzmann suppressed since they are created in a phase where they are more massive. Also, to interact with the wall, they need to be faster than $v_w$ to catch up with it, which only a small fraction of the particles can do (intuitively, we expect this fraction to go from $1/2$ at $v_w=0$ to 0 at $v_w=1$). It is therefore a good approximation to neglect $\delta f_-$ in the pressure calculation and set $\delta f\approx \delta f_+$. This fact can be seen explicitly in Fig.~\ref{fig:ballisticPressures} which shows the contribution of the different modes to the ballistic pressure. In both the strong and weak PTs, the $t^-$-modes (which correspond to $\delta f_-$) give the smallest contribution, and they are completely negligible in the strong PT.

From Eqs.~(\ref{eq:boltzmannSolution}) and (\ref{eq:lambdaVariation}), one finally gets $\delta f>0$ and $\partial_\lambda \delta f < 0$. From Eq.~\eqref{eq:Pfriction}, one has
\begin{align}
    \P_{\rm friction}=\P_{\rm LTE} - \sum_i\int_{-\delta}^\delta \d z\, \underbrace{ \frac{\d m^2_i(\phi(z))}{\d z}  }_{<0}\underbrace{\left(\int \frac{\d^3{\bf p}}{(2\pi)^32E_i}\,\delta f_i\right)}_{> 0} \,,
\end{align}
where $\P_{\rm LTE}$ is obtained by substituting $f_i=f_i^{\rm eq}$ into Eq.~\eqref{eq:Pfriction}.
$\delta f_i>0$ means that out-of-equilibrium effects add more friction compared with the LTE limit. While $\partial_\lambda \delta f_i <0$ means that the minimum is reached when $\lambda\to\infty$ (LTE) and the maximum pressure is reached when $\lambda=0$ (ballistic limit).


\section{Comparison between LTE and ballistic results}
\label{sec:numerics}

As the LTE and ballistic approximations respectively offer an upper and lower bound on the true wall velocity, it can be useful to compare these two approximations to find constraints on the wall velocity without having to solve a complicated system of Boltzmann equations and scalar EoM. Not only are these equations challenging and computationally expensive to solve, they also suffer from large theoretical uncertainties coming from the collision rates $\Gamma$ that appear in the Boltzmann equations. These collision rates are typically computed to leading-log accuracy, which is rather imprecise and can lead to errors of $\mathcal{O}(1)$ \cite{Moore:1995si,Cline:2021iff,Laurent:2022jrs,Kozaczuk:2015owa}. 

The LTE and ballistic approximations allow us to avoid this problem since they correspond to the limits $\Gamma\to\infty$ and $\Gamma\to 0$, respectively. Thus, even if $\Gamma$ is completely unknown, one can still obtain useful information on $v_w$ with these approximations in the form of an interval in which it must reside.

Unfortunately, making a general comparison between LTE and ballistic is not a straightforward task as these two approximations depend on a different set of variables. In general, this mapping from one approximation to the other cannot be made in a model-independent way. It either necessitates specifying a model, in which case the mapping can be done exactly, or making approximations which allows one to make more general conclusions that should still hold at least qualitatively. We will explore both options in the following.

\subsection{Model-independent analysis}
\label{sec:model-independent-ballistic}

To obtain a more general and qualitative understanding of how the LTE and ballistic approximation compare to one another, we now study a set of simplified models. We will assume that the plasma contains ${\tilde{g}_{\star}}$ massless degrees of freedom and ${g_{*,\phi}}$ degrees of freedom with a mass of $m_+$ and $m_-$ outside and inside the bubble, respectively.
The pressure in this model can therefore be expressed as 
\begin{align}
\label{eq:EOS}
    p_{s/b}(T) = \frac{\pi^2 {\tilde{g}_{\star}}}{90}T^4 + {g_{\star,\phi}}\, p_{M,s/b}(T) - \epsilon_\pm\,,
\end{align}
where $\epsilon_\pm$ is the vacuum contribution (which is approximated to be independent of $T$), and $p_{M,s/b}$ is the thermal pressure of a single degree of freedom with mass $m_\pm$. To simplify the discussion, we will assume that $p_{M,s/b}$ is given by the one-loop Maxwell-Boltzmann pressure, which is given by
\begin{align}
    p_{M,s/b}(T) =\frac{\pi^4}{90} \frac{m_\pm^2 T^2}{2\pi^2} K_2\left(\frac{m_\pm}{T}\right)\,,
\end{align}
where $K_n(x)$ is the modified Bessel function of the second kind. Above, we have added by hand an extra factor $\pi^4/90\approx 1.08$ so that $p_M$ reproduces the value of the Bose-Einstein distribution
for $m_{\pm}=0$. This way, $g_{\star,\phi}$ can be taken to be the effective number of degrees of freedom coupling with the scalar $\phi$,
\begin{align}
    {g_{\star,\phi}}=\sum_{i_B}g_{i_B,\phi}+\sum_{i_F} \frac{7}{8} g_{i_F,\phi} \,,
\end{align}
where $g_{i_B,\phi}$ is the number of internal degrees of freedom of boson $i_B$, and $g_{i_F,\phi}$ of fermion $i_F$. {When $m_+=0$, the total number of relativistic degrees of freedom in the symmetric phase is $\tilde{g}_\star+g_{\star,\phi}\equiv g_\star$.}

It follows directly that the enthalpy and energy densities are given by
\begin{subequations}
\begin{align}
    \omega_{s/b}(T) &=  \frac{2\pi^2 {\tilde{g}_\star}}{45}T^4+{g_{\star,\phi}}\frac{\pi^4}{90}\frac{m_\pm^3 T}{2\pi^2}K_3\left(\frac{m_\pm}{T}\right)\,,\\
    e_{s/b}(T) &= \frac{\pi^2 {\tilde{g}_\star}}{30}T^4 + {g_{\star,\phi}}{\frac{\pi^4}{90}}\frac{m_\pm^2 T}{2\pi^2}\left[ m_\pm K_3\left(\frac{m_\pm}{T}\right) - T K_2\left(\frac{m_\pm}{T}\right)\right] + \epsilon_\pm\,.
\end{align}
\end{subequations}

At first sight, it could appear that the EoS \eqref{eq:EOS} depends on six parameters (${\tilde{g}_\star}$, ${g_{\star,\phi}}$, $m_\pm$ and $\epsilon_\pm$). However, one can see that the variables ${\tilde{g}_\star}$, ${g_{\star,\phi}}$ and $\epsilon_\pm$ only appear in the combinations 
\begin{align}
    b = \frac{{g_{\star,\phi}}}{{\tilde{g}_\star}}\,,\qquad\qquad
    \epsilon = \frac{\epsilon_+ - \epsilon_-}{{\tilde{g}_\star}},
\end{align}
in all the matching equations. This effectively lowers the number of independent variables in this model to four.

Finally, to simplify the LTE calculation, one can map the EoS \eqref{eq:EOS} to the template model (see Appendix \ref{app:template}) by computing the thermodynamic quantities $\alpha(T_n)$, $c_{s/b}(T_n)$ and $\Psi(T_n)$ with Eqs.~(\ref{eq:alpha}), (\ref{eq:thermo-quantities-from-p}) and (\ref{eq:Psi}). This typically gives a reasonable approximation and allows one to use the method presented in Ref.~\cite{Ai:2023see} to compute $v_w$. However, to guarantee accurate results, we will keep the full EoS \eqref{eq:EOS} in what follows and use the {\tt WallGo} package \cite{Ekstedt:2024fyq} to find the LTE wall velocity (based on hydrodynamics \cite{Ai:2021kak}).

\subsubsection{Results}

We now move to comparing the wall velocity obtained within this model using the ballistic and LTE limits. As argued previously, these limits give respectively a lower and upper bound on the possible wall velocities and can therefore contain helpful information. We will be particularly interested in the difference between these two predictions. If the difference is large (or if they predict different types of solutions), the true wall velocity remains poorly constrained and it is required to solve a set of Boltzmann equations to obtain a better estimate. On the other hand, if the difference is small, the collision operator has very little effect on the wall velocity and it may be unnecessary to solve the Boltzmann equations. 

In what follows, it will be convenient to describe the strength of the PT with a slightly different definition of $\alpha$ than the one defined in terms of the pseudo trace $\bar\theta$ (see Eq.\ (\ref{eq:alpha})). Instead, we will use a definition based on the pressure difference:
\begin{align}
\label{eq:alphap}
    \alpha_p(T) = -(1+1/c_b^2)\left(\frac{p_s(T)-p_b(T)}{3\omega_s(T)}\right) = \alpha(T) - \frac{1-\Psi(T)}{3}.
\end{align}
The two definitions are approximately equal for strong PTs, but slightly disagree for weaker ones. In particular, by definition, $\alpha_p$ vanishes when $\Delta V=0$, so it can always take values from 0 to $\infty$, while $\alpha$ cannot be smaller than $(1-\Psi)/3$. Furthermore, the wall velocity always goes to 0 in the limit $\alpha_p\to 0$. These properties make it easier to interpret and present results expressed in terms of $\alpha_p$ rather than $\alpha$ (one can of course always go easily from one to the other if needed using Eq.\ (\ref{eq:alphap})).

\paragraph{Deflagrations/hybrids}

\begin{figure}
    \centering
    \vspace{-1.3cm}
    \footnotesize{$m_+=0$ and $m_-=T_n$ \hspace{0.25\linewidth} $m_+=0$ and $m_-=\infty$} \\
    \includegraphics[width=0.5\linewidth]{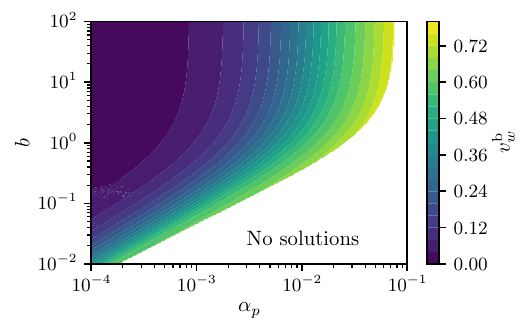}\hspace{0.0\linewidth}\includegraphics[width=0.5\linewidth]{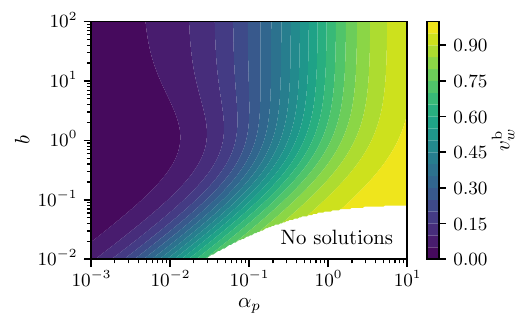}\\[-2.5mm]
    \includegraphics[width=0.5\linewidth]{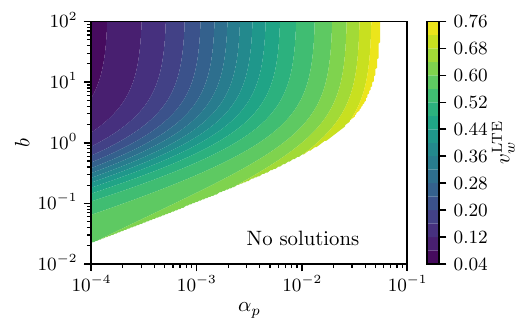}\hspace{0.0\linewidth}\includegraphics[width=0.5\linewidth]{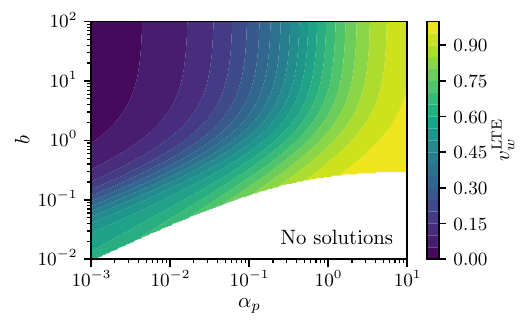}\\[-2.5mm]
    \includegraphics[width=0.5\linewidth]{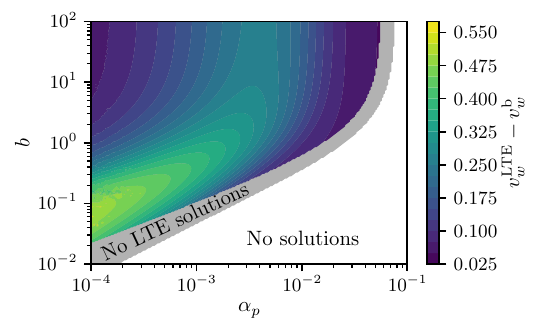}\hspace{0.0\linewidth}\includegraphics[width=0.5\linewidth]{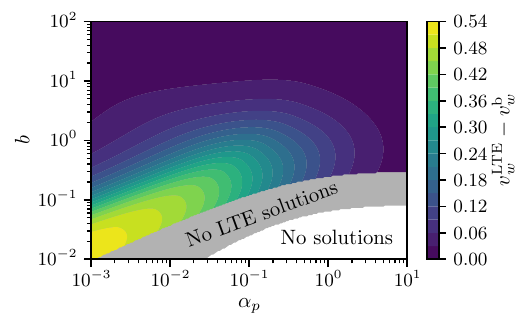}\\[-2.5mm]
    \includegraphics[width=0.5\linewidth]{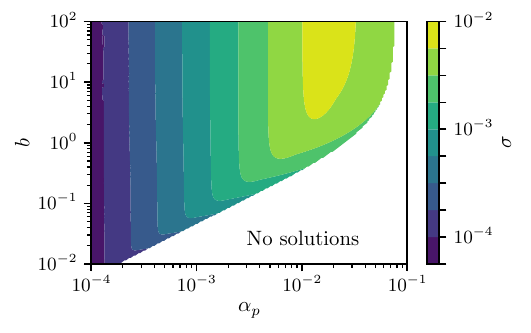}\hspace{0.0\linewidth}\includegraphics[width=0.5\linewidth]{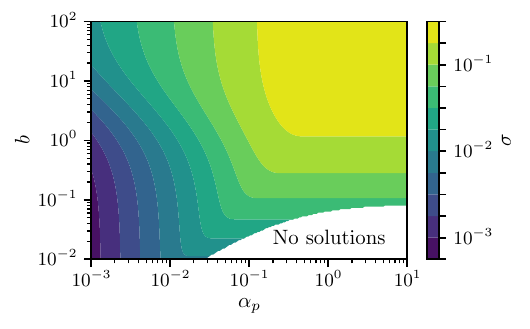}\\[-6mm]
    \caption{\footnotesize{Contour plots of scans varying the parameters $\alpha_p$ and $b$ with fixed $m_+$ and $m_-$ for deflagration and hybrid solutions only. They show (from top to bottom) the ballistic solution $v_w^{\rm b}$, the LTE solution $v_w^{\rm LTE}$, the difference between the ballistic and LTE solutions $v_w^{\rm LTE}-v_w^{\rm b}$ and the entropy fraction generated by the ballistic solution $\sigma$. The left side is with $m_+=0$ and $m_-=T_n$, and the right side with $m_+=0$ and $m_-=\infty$ (the large mass limit). }}
    \label{fig:alphaBScan}
\end{figure}

\begin{figure}
    \centering
    \vspace{-1.3cm}
    \footnotesize{$\alpha_p=0.01$ and $b=0.1$ \hspace{0.3\linewidth} $\alpha_p=1$ and $b=1$} \\
    \includegraphics[width=0.5\linewidth]{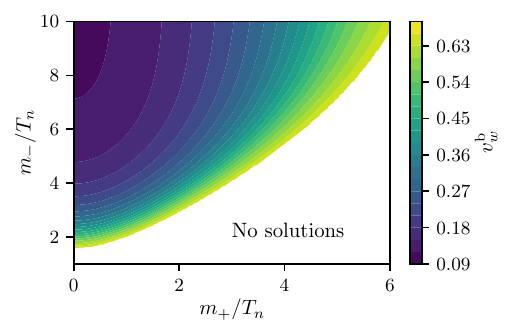}\hspace{0.0\linewidth}\includegraphics[width=0.5\linewidth]{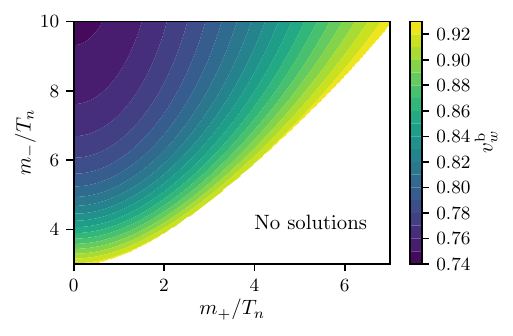}\\[-2.5mm]
    \includegraphics[width=0.5\linewidth]{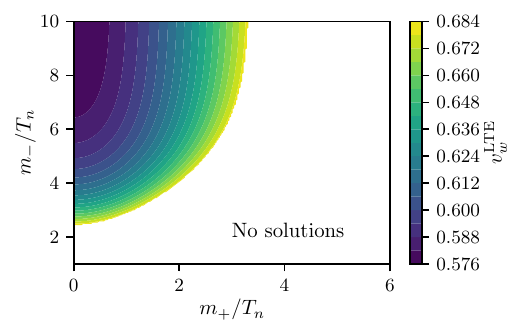}\hspace{0.0\linewidth}\includegraphics[width=0.5\linewidth]{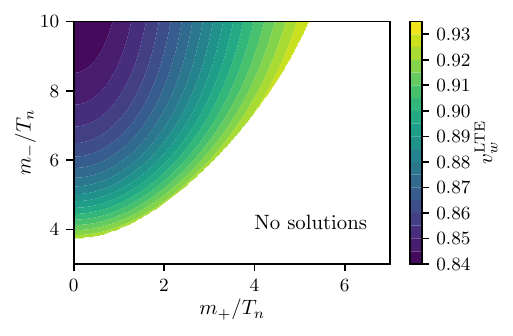}\\[-2.5mm]
    \includegraphics[width=0.5\linewidth]{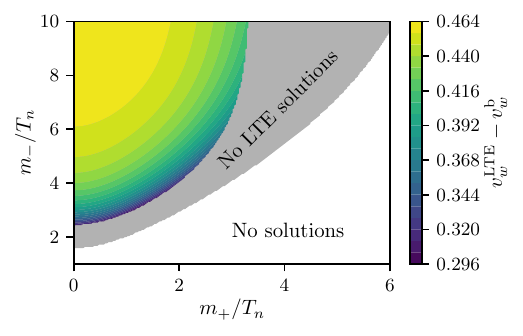}\hspace{0.0\linewidth}\includegraphics[width=0.5\linewidth]{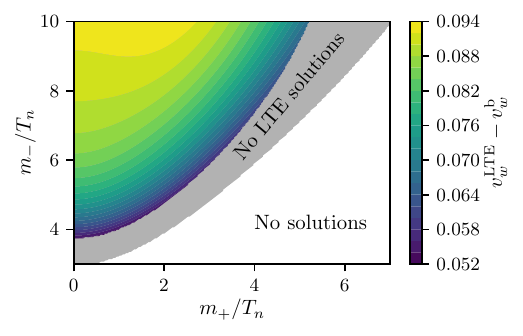}\\[-2.5mm]
    \includegraphics[width=0.5\linewidth]{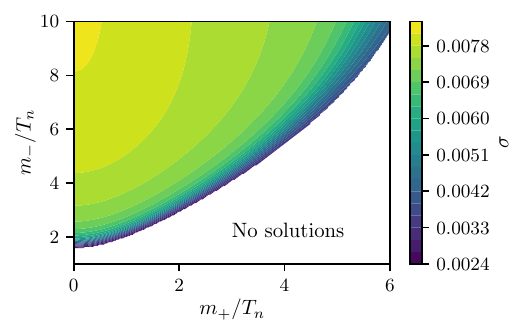}\hspace{0.0\linewidth}\includegraphics[width=0.5\linewidth]{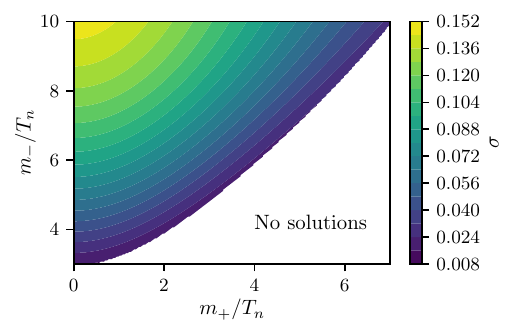}\\[-6mm]
    \caption{\footnotesize{Contour plots of scans varying the parameters $m_+$ and $m_-$ with fixed $\alpha_p$ and $b$ for deflagration and hybrid solutions only. They show the same quantities as in Fig.\ \ref{fig:alphaBScan}. The left side is for a rather weak PT with $\alpha_p=0.01$ and $b=0.1$, while the right side is for a stronger PT with $\alpha_p=1$ and $b=1$.}}
    \label{fig:m1m2scan}
\end{figure}

To explore the properties of deflagration and hybrid solutions we perform a scan varying $\alpha_p$ and $b$ while keeping $m_\pm$ fixed. The result of this scan is shown in Fig.\ \ref{fig:alphaBScan}, with $m_+=0$ and $m_-=T_n$ on the left panels and $m_+=0$ and $m_-=\infty$ (more formally $m_-\gg\gamma_w T_n$) on the right panels. The former is meant to represent a typical electroweak PT (with the Higgs acquiring a VEV $v\sim T_n$), while the latter corresponds to the large mass limit (LML) which has interesting properties that will be discussed in the next subsection. For these two cases, one has $\Psi(T_n)\approx (1+0.89 b)/(1 +b)$, $\Psi(T_n)=1/(1 +b)$, respectively. This means that the parameter $b$ has a very different implication in these two cases. For the former $\Psi$ can never be smaller than 0.89, while for the latter, it can be as small as 0.

The first two rows of Fig.\ \ref{fig:alphaBScan} show the wall velocity computed in the ballistic limit $v_w^{\rm b}$ and in LTE $v_w^{\rm LTE}$. In all cases, we can see white regions where no deflagration or hybrid solutions were found. In these regions, the PTs are simply too strong and the plasma too weakly coupled to create substantial friction on the wall. The wall is therefore free to accelerate beyond the Jouguet velocity and become a detonation or possibly a runaway solution. The part of parameter space where a solution can be found is always larger when the ballistic approximation is used rather than LTE, since the friction in the first case is higher and can therefore stop the wall more easily. Note that the same conclusion does not hold if one uses the B\"{o}deker-Moore 1-to-1 thermal friction~\eqref{eq:frictonBM}. This means that the asymptotic value of the pressure in the $\gamma_w\rightarrow \infty$ (when only the 1-to-1 processes are considered) is not necessarily larger than the pressure peak at the Jouguet velocity~\cite{Ai:2024shx}. 

Note that the $m_-=T_n$ and $m_-=\infty$ cases behave in a very different way. In the former, there is a maximal value of $\alpha_p$ beyond which no deflagration or hybrid solutions are possible. In the second case, if $b$ is high enough, there will always be such a solution no matter how strong the PT is. This fact was first observed in Ref.~\cite{Ai:2023see} for LTE, and we can observe here that it can also happen in the ballistic limit. 
Note however that the ballistic wall velocity can only be interpreted as an upper bound as long as the size of the shock wave is larger than $ L_{\rm MFP}$. 

The third row of Fig.\ \ref{fig:alphaBScan} shows the difference between $v_w^{\rm LTE}$ and $v_w^{\rm b}$. An important observation is that, in most of the parameter space, both limits agree on the type of solution. Effectively, they both agree that a deflagration/hybrid solution exists in the colored region and that no solution exists in the white region (which means the wall must be a detonation or runaway). They only disagree in the narrow grey band where a ballistic solution can be found but no LTE one. This implies that, away from this grey band, it is possible to determine with certainty the solution type of a general model: deflagration/hybrid if a solution exists in LTE, and detonation if no solution exists in ballistic.

If there is a solution in LTE, then the true wall velocity $v_w$ is in the interval $[v_w^{\rm b},v_w^{\rm LTE}]$. A simple estimate of $v_w$ can then be obtained with, e.g., $v_w^{\rm mean}=(v_w^{\rm b}+v_w^{\rm LTE})/2$. As can be seen in Fig.\ \ref{fig:alphaBScan}, the difference between $v_w^{\rm b}$ and $v_w^{\rm LTE}$ can be quite large in some parts of the parameter space, and $v_w^{\rm mean}$ could be off by at most $\sim 0.3$. If that is the case, solving the Boltzmann equations could be necessary if more precision is needed. There are two regimes where $v_w^{\rm b}$ and $v_w^{\rm LTE}$ are approximately equal. The first one happens when $\Psi\approx 0$ (or $b \gg 1$ and $m_- - m_+\gg T_n$) where, as argued in {Section \ref{sec:LTE_as_bound_template}}, $\Psi_{\rm eff}\approx \Psi$ and the entropy production has no effect on the wall velocity. $v_w^{\rm mean}$ also becomes exact when $\alpha_n\gg 1/3$ simply because both $v_w^{\rm b}$ and $v_w^{\rm LTE}$ approach 1 for strong PTs. Both of these two cases can be seen on the right side of Fig.\ \ref{fig:alphaBScan}. If a model is in one of these two situations (which often happens in confining PTs or in models with conformal invariance), solving the Boltzmann equations may become unnecessary as $v_w^{\rm b}$ and $v_w^{\rm LTE}$ agree very well.

Finally, the fourth row of Fig.\ \ref{fig:alphaBScan} shows the entropy production fraction $\sigma$ (Eq.\ (\ref{eq:sigma})) obtained in the ballistic approximation. This quantity can be useful to quantify the extent to which the plasma can be out-of-equilibrium. It can also be used to compute $\Psi_{\rm eff}$ which allows one to describe mathematically the model using the LTE formalism, even if the plasma has deviations from equilibrium. Furthermore, $\sigma$ always vanishes in LTE, while it is maximized in the ballistic limit, so the true entropy produced must always be smaller than the value shown here.

We show different scans in Fig.\ \ref{fig:m1m2scan} where $\alpha_p$ and $b$ are held fixed and $m_\pm$ are varied. Again, there are two different scans: one with $\alpha_p=0.01$ and $b=0.1$ which are typical values obtained in the singlet scalar extension of the SM, and the other one with $\alpha_p=1$ and $b=1$ which represents a stronger PT. One can observe that the wall velocity is minimized in the LML (the upper left corner), where the mass variation is maximal. This is not surprising as the large mass variation creates a large friction that slows down the wall more efficiently. Furthermore, it can be seen that having a deflagration or hybrid solution requires having a relatively small value of $m_+$. If it is too large, the particles are completely Boltzmann suppressed in front of the wall and they cannot generate any pressure, even if $m_-\gg m_+$. We note, however, that both $v_w^{\rm LTE}-v_w^{\rm b}$ and $\sigma$ are maximized in the LML.

\paragraph{Detonations}

\begin{figure}
    \centering
    \footnotesize{$m_+=0$ and $m_-=T_n$ \hspace{0.3\linewidth} $m_+=0$ and $m_-=5T_n$} \\
    \includegraphics[width=0.5\linewidth]{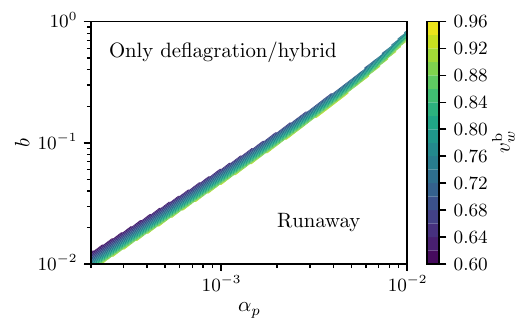}\includegraphics[width=0.5\linewidth]{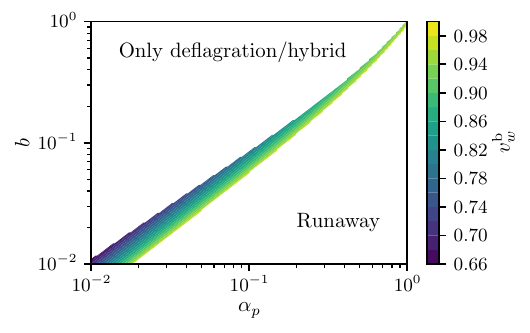}\\[-2mm]
    \includegraphics[width=0.5\linewidth]{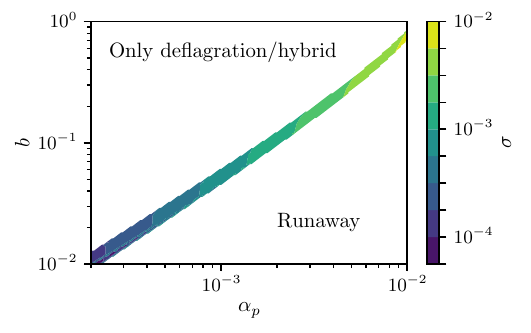}\includegraphics[width=0.5\linewidth]{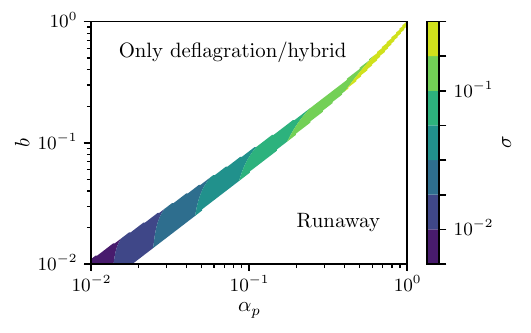}\\[-4mm]
    \caption{\footnotesize{Contour plots of scans varying the parameters $\alpha_p$ and $b$ with fixed $m_+$ and $m_-$ for detonation solutions only. They show the ballistic solution $v_w^{\rm b}$ and the entropy fraction generated $\sigma$. The left side is with $m_+=0$ and $m_-=T_n$, and the right side with $m_+=0$ and $m_-=5T_n$.}}
    \label{fig:alphaBScanDeton}
\end{figure}

We now move to describe the detonation solutions. In contrast to deflagration and hybrid solutions, it was shown in Ref.\ \cite{Ai:2023see} that stable detonations do not occur in LTE. This is because in LTE the net pressure $\P_{\rm friction}(v_w)-\P_{\rm driving}$, is a monotonically decreasing function of $v_w$ after the Jouguet velocity. Even though there is a solution $\P_{\rm friction}(v_w)-\P_{\rm driving}=0$ for $v_w> v_J$, it cannot be stable.  On the other hand, we will see shortly that detonations do occur in the ballistic limit as now the net pressure is not monotonically decreasing above $v_J$ (see Fig.~\ref{fig:pressurexSM} below for an example).  There are even parameter points with several detonation solutions, but for simplicity, we will only consider the solution with the smallest wall velocity in what follows.

\begin{figure}
    \centering
    {\footnotesize $\alpha_p=0.01$ and $b=0.1$} \hspace{0.3\linewidth} {\footnotesize $\alpha_p=1$ and $b=1$} \\
    \includegraphics[width=0.5\linewidth]{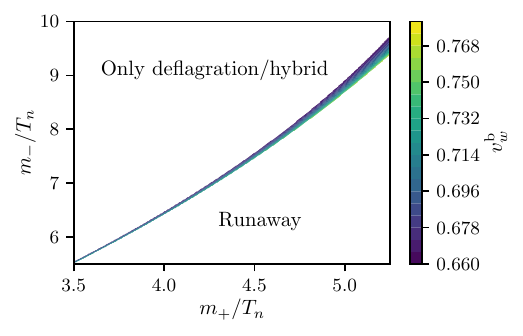}\includegraphics[width=0.5\linewidth]{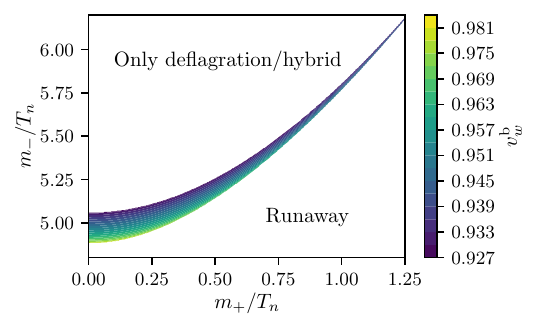}\\[-2mm]
    \includegraphics[width=0.5\linewidth]{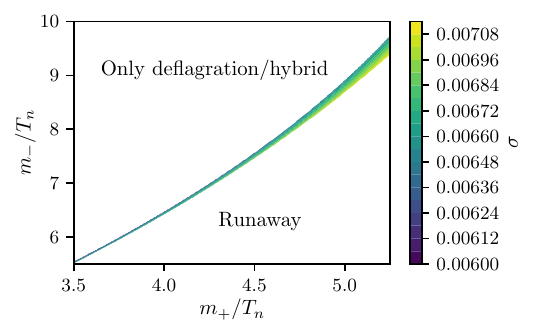}\includegraphics[width=0.5\linewidth]{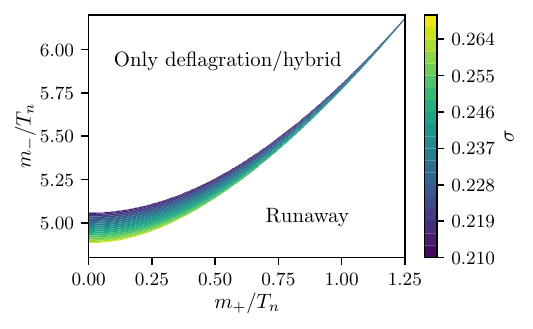}\\[-4mm]
    \caption{\footnotesize{Same as in Fig.\ \ref{fig:alphaBScanDeton} but varying $m_\pm$ with $\alpha_p=0.01$ and $b=0.1$ (left) and $\alpha_p=b=1$ (right).}}
    \label{fig:m1m2scanDeton}
\end{figure}

We show in Figs.\ \ref{fig:alphaBScanDeton} and \ref{fig:m1m2scanDeton} the same scans as for the deflagration and hybrid branch. It is quite clear from these scans that finding a stable static detonation solution requires some amount of fine-tuning. If the PT is too weak, the driving force on the wall will simply be too small to overcome the friction in the detonation regime. In that case, the net pressure on the wall is always positive for $v_w\in [v_J, 1)$ and the only possible solutions are deflagrations or hybrids. 
On the other hand, in the detonation regime the friction generated by the plasma cannot be arbitrarily large (see the Bödeker-Moore limit in Sec.~\ref{sec:ballistic}), so if the PT is too strong, the net pressure will always be negative for $v_w> v_J$. In that case, although no static detonations can be found, the wall can become a runaway as the negative pressure will continue pushing the wall forward until it reaches an ultrarelativistic speed.

\subsubsection{Large mass limit}

An interesting case happens when the massive particles start nearly massless in the symmetric phase and end up very heavy in the broken phase, which corresponds to the limit $m_+ \to 0$ and $m_- \to \infty$ (or more specifically, $m_-\gg \gamma_+ T_+$). This type of situation can naturally occur in models where a scalar field acquires a very large VEV (e.g.\ in conformal models) or in confining PTs.\footnote{At first glance, the ballistic treatment seems not particularly useful in the confining case. Particles cannot simply be tracked from the deconfined phase into the confining phase, since both phases contain completely different degrees of freedom. In the limit $m_- \to \infty$ however, only particles in the deconfined phase are relevant, and the ballistic approximation can be used to describe those.}
In this large mass limit (LML), the plasma only contains massless particles everywhere, as the heavy particles are completely Boltzmann suppressed in the broken phase, which thus only contains the massless species. As we will see, this fact drastically simplifies the analysis.

The LML is not only simpler to analyze, it also provides an interesting limit. For a fixed value of $\alpha_p$ (or $\alpha_n$) and $b$, it becomes clear from  the second row of Fig.\ \ref{fig:m1m2scan} that the LML gives the smallest value of the wall velocity. The LML thus provides us with a lower bound on the ballistic wall velocity, which was already the lower bound on the wall velocity. Put in another way, the LML gives the lowest lower bound on the wall velocity among all particle physics models with a fixed $\alpha_p$ and $b$.

Let us now demonstrate how the analysis simplifies in the LML. A first simplification occurs for the equation of state of the LML. Since the plasma is completely radiation-dominated, we always have $c_{s/b}^2=1/3$. This implies that the plasma can be represented perfectly by the bag EoS. 
As mentioned before, $\Psi = 1/(1+b)$ for this case. Notice that these thermodynamic quantities are all temperature-independent, which is not generally the case. 

The second simplification happens in the calculation of the ballistic pressure. Because $m_- \to \infty$, it can be seen in Eqs.\ (\ref{eq:P-ballistic-momentum-final-ex}) that only the contribution from reflections $\P^r$ is nonvanishing. Furthermore, the integral can be performed analytically which gives a total friction force on the wall,
\begin{align}
\label{eq:frictionLargeMass}
    \P_{\rm friction}(T_+, v_+) = \frac{\pi^2 {g_{\star,\phi}} T_+^4}{90}(1+v_+)^3\gamma_+^2 = \frac{1-\Psi}{4}\,\omega_+(T_+)(1+v_+)^3\gamma_+^2\,.
\end{align}
Notice that this pressure asymptotically grows like $\gamma_+^2$ when $v_+\to 1$, and can therefore be arbitrarily large. Therefore, there are no runaway solutions in the LML. This seems to be in contradiction with the Bödeker-Moore limit~\cite{Bodeker:2009qy} mentioned in Section \ref{sec:ballistic}, which predicts a finite pressure in the $v_+\to 1$ limit. This discrepancy is a consequence of the order in which the two limits are taken. By taking the LML first, we implicitly assume $m_-/T_+\gg \gamma_+$, which gives the friction (\ref{eq:frictionLargeMass}). On the other hand, Bödeker-Moore assumes $m_-/T_+\ll \gamma_+$, which gives Eq.\ (\ref{eq:frictonBM}). In realistic models, $m_-$ can be large, but it is finite. Thus, the LML cannot be valid up to $v_+=1$ and for a large enough wall velocity, the Bödeker-Moore limit will become the correct description. We will however not consider such complications and focus on the pure LML in what follows.

By requiring that $\P_{\rm driving}=\P_{\rm friction}$, one can derive a new condition valid in the ballistic LML:
\begin{align}
\label{eq:largeMassMatching}
    \alpha(T_+) = \frac{1-\Psi}{3}(1+v_+)^3\gamma_+^2\,.
\end{align}
Interestingly, this condition only depends on the plasma velocity and temperature measured in front of the wall. 

\paragraph{Detonations} For detonation solutions which have $v_+ = v_w$ and $T_+ =T_n$, the wall velocity can be determined directly to be
\begin{align}
\label{eq:largeMassDet}
    v_w^{\rm det} = \frac{\sqrt{3\alpha_n [8(1-\Psi)+3\alpha_n]}-3\alpha_n}{2(1-\Psi)}-1\,.
\end{align}
For this solution to be valid, it must be larger than the Jouguet velocity \cite{Ai:2023see}
\begin{align}
    v_J = \frac{1}{\sqrt{3}}\left(\frac{1+\sqrt{\alpha_n(2+3\alpha_n)}}{1+\alpha_n}\right)\,,
\end{align}
otherwise, a shock wave would form in front of the wall and the solution would be a deflagration or hybrid wall, not a detonation. This implies that to have a detonation solution in the LML, one must satisfy
\begin{align}
\label{eq:condition_Psidet}
    \Psi_{\rm det} > 1-\frac{3\alpha_n(1-v_J)}{(1+v_J)^2}\,.
\end{align}
In particular, one can show that this bound is minimal when $\alpha_n\to\infty$, in which case the bound becomes $\Psi_{\rm det} > (2+\sqrt{3})/4\approx 0.933$. Therefore, in the ballistic LML, no detonation solution can exist if $\Psi$ is smaller than 0.933 no matter what $\alpha_n$ is.  For $\Psi< 0.933$, the wall velocity 
would be too slow to prevent a shock wave from forming, and the solution would eventually become a hybrid or deflagration.

\begin{figure}[t]
    \centering
\includegraphics[width=0.6\linewidth]{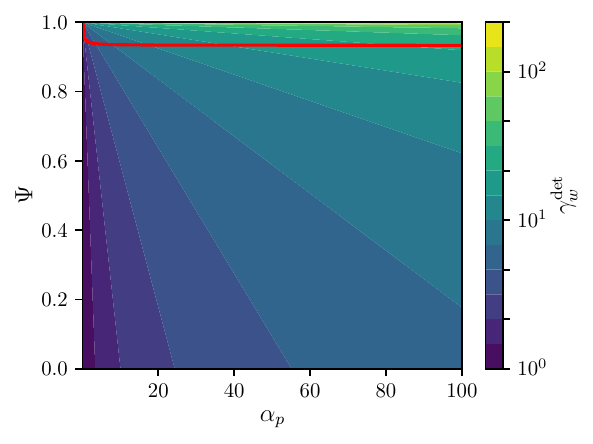}
    \caption{Contour plot of $\gamma_w^{\rm det}(\alpha_p,\Psi)$ for the velocity given in Eq.~\eqref{eq:largeMassDet}. The region below the red line violates the condition~\eqref{eq:condition_Psidet}. If further $m_-/T_n$ is much larger than $\gamma_w^{\rm det}$, then there is no detonation solution for the parameter region below the red line.
    }
    \label{fig:gammaw}
\end{figure}

As mentioned earlier, in realistic situations $m_-$ cannot be infinite. Therefore, the above conclusion is valid only for 
\begin{align}
    \frac{m_-}{T_n} \gg \gamma_w^{\rm det}\,,
\end{align}
which gives an additional condition for $m_-/T_n$ for given $\alpha_n$ and $\Psi$. In Fig.~\ref{fig:gammaw}, we show $\gamma_w^{\rm det}$ together with the condition~\eqref{eq:condition_Psidet}. It can be seen that for the plotted $\alpha_p\lsim  100$, a value of $m_-/T_n$ of order $\O(10)-\O(100)$  would justify the conclusion that there is no detonation solution when $\Psi \lsim 0.933$. 

\paragraph{Deflagrations/hybrids}
The situation for deflagration and hybrid solutions is more complicated since they have a shock wave propagating in front of the wall, so one cannot directly relate $v_+$ and $T_+$ to $v_w$ and $T_n$. Nevertheless, one can use the new condition (\ref{eq:largeMassMatching}) to compute the entropy production fraction $\sigma$ defined in Eq.\ (\ref{eq:sigma}). As described in Section \ref{sec:LTE_as_bound_template}, this will allow us to treat the system like it is in LTE, but with an effective value of $\Psi$
\begin{align}
\label{eq:Psi_eff}
    \Psi_{\rm eff} = \frac{\Psi}{(1+\sigma)^4}\,.
\end{align}
This thus gives us a direct way to compare LTE to ballistic, since they can be treated mathematically identically.

First, one can eliminate $T_-$ from the definition of $\sigma$ by substituting Eq.\ (\ref{eq:conditionA}) into (\ref{eq:sigma}) and (\ref{eq:entropycreation}). By assuming that the plasma is described by the bag EoS (which is always the case in the LML), one can show that
\begin{align}
\label{eq:sigmaSimplified}
    \sigma = \left(\frac{\gamma_+^2 v_-}{\gamma_-^2 v_+}\Psi\right)^{1/4}-1\,.
\end{align}
One can then substitute Eq.\ (\ref{eq:largeMassMatching}) into (\ref{eq:match1}) to obtain a matching equation independent of $T_-$ and $T_+$. It can be expressed in terms of the cubic equation
\begin{align}
\label{eq:cubicMatching}
    0=(1-\Psi)v_- v_+^3-3\Psi v_- v_+^2 + [1+3v_-(1+v_- - \Psi)]v_+ - \Psi v_-\,.
\end{align}
In principle, one could use Cardano's formula to solve this cubic equation for $v_+$ and substitute the result into Eq.~\eqref{eq:sigmaSimplified} to express $\sigma$ completely in terms of $\Psi$ and $v_w$ (remember that $v_-=v_w$ for deflagrations and $v_-=c_b=1/\sqrt{3}$ for hybrids). However, the resulting expression is quite complicated which makes it challenging to analyse and understand its qualitative behavior.

What will be much more instructive is to study different limiting cases. In particular, we will be interested in the small $v_-$, small $\Psi$ and small $1-\Psi$ limits. In these limits, the solution of Eq.\ (\ref{eq:cubicMatching}) is
\begin{align}
    v_+ = 
    \begin{cases} \frac{v_- \Psi}{1+3v_- (1+v_- -\Psi)}, & v_-\ll 1\ \mathrm{or}\ \Psi\ll 1\,, \\ 
    v_- - \frac{v_- (1+v_-)^3 (1-\Psi)}{1-3v_-^2}, & 1-\Psi\ll \frac{1}{\sqrt{3}}-v_-\,.
    \end{cases}
\end{align}
Substituting these into Eq. (\ref{eq:sigmaSimplified}), one obtains the entropy production fraction
\begin{equation}
\label{eq:sigmaLimits}
    \sigma = 
    \begin{cases}
        \frac{(3+v_-)}{4(1-v_-)}v_- (1-\Psi), & v_- \ll 1 \ \mathrm{or}\ 1-\Psi \ll 1\,,\\
        \frac{f(v_-)}{\sqrt{\gamma_-}}-1-\frac{3v_- \Psi}{4f^3(v_-)\sqrt{\gamma_-}}, & \Psi\ll 1\,,
    \end{cases}
\end{equation}
with $f(v_-)=[1+3v_-(1+v_-)]^{1/4}\approx 1+0.675 v_-$.

There are a few interesting consequences of these limiting behaviours. First, $\sigma$ always vanishes when $v_-=0$ or $\Psi=1$. Also, it is an increasing function of $v_-$ and a decreasing function of $\Psi$. This implies that for deflagration and hybrid solutions, $\sigma$ is maximized when $\Psi=0$ and $v_-=1/\sqrt{3}$, at which point it takes the value $\sigma_{\rm max}=f(1/\sqrt{3})(2/3)^{1/4}-1\approx 0.256$. Because the pressure in the ballistic LML is the absolute maximum a PT with a fixed $\alpha_p$ and $b$ can have, $\sigma_{\rm max}$ is actually the largest entropy fraction a PT with deflagration or hybrid expansion can produce, no matter what the model is and whether it is best described by LTE, ballistic, or anything in between. However, note that for detonations, it can be shown that $\sigma$ grows asymptotically like $\alpha^{1/4}$ and can therefore be arbitrarily large. 

Starting from Eq.\ (\ref{eq:sigmaLimits}), it is quite easy to find a numerical fit that approximates $\sigma$ well for every value of $\Psi$ and $v_-$. For example, a good fit can be obtained with the simple function
\begin{align}
    \sigma_{\rm fit}=\left(\frac{f(v_-)}{\sqrt{\gamma_-}}-1\right)\frac{1-\Psi}{1-A v_- \Psi}\,,
\end{align}
with $A\approx 0.91$ giving the best fit with a maximal error of less than 1\% for $\Psi\in[0,1]$ and $v_-\in [0,1/\sqrt{3}]$. Furthermore, this fit is exact when $\Psi=0,\, 1$ or $v_-=0$. Once we have $\sigma_{\rm fit}$, we can substitute it into Eq.~\eqref{eq:Psi_eff} to get $\Psi_{\rm eff}(T_n)$ for a given $\Psi(T_n)$. Then substituting $\Psi_{\rm eff}(T_n)$ into the fit formula for the LTE wall velocity given in Ref.~\cite{Ai:2023see}, we obtain the resulting wall velocity. Note that we are only using the correspondence mentioned in Sec.~\ref{sec:LTE_as_bound_template} and are still deriving the ballistic wall velocity here.

\subsection{Example model: the Standard Model coupled to a gauge singlet}
\label{sec:example-model}

Let us now put our approximations to the test in a concrete example model. We choose the $\mathbb Z_2$-symmetric xSM, where the Standard Model Higgs $h$ is coupled to a new gauge singlet $s$ (see e.g. Refs.~\cite{Espinosa:1993bs, Barger:2007im, Espinosa:2011ax, Kozaczuk:2015owa, Laurent:2022jrs}), with tree level potential
\begin{align}
    V(h,s)=-\frac{\mu_h^2}{2}h^2+\frac{\lambda_h}{4}h^4+\frac{1}{2}\left(m_s^2-\frac{\lambda_{hs}v^2}{2}\right)s^2+\frac{\lambda_s}{4}s^4+\frac{\lambda_{hs}}{4}h^2 s^2\,,
\end{align}
where $v=\mu_h/\sqrt{\lambda_h}=246\,{\rm GeV}$ is the VEV of the SM Higgs $h$. $m_s$ is the mass of the scalar singlet $s$ at the vacuum $(v,0)$ given by $\langle h\rangle=v, \langle s\rangle =0$. We consider a region of parameter space where the PT proceeds in two steps: first, the singlet obtains a VEV, and in the second step, the Higgs obtains a VEV and the singlet goes back to zero. We use Benchmark 1 defined in Ref.~\cite{Ekstedt:2024fyq}, with $m_s = 120 \, {\rm GeV}$, $\lambda_{hs}$ and $\lambda_s = 1$ and focus on the second step of the PT for two choices of the nucleation temperature $T_n = 90, 100 \, {\rm GeV}$.

We implement the one-loop effective potential as in Ref.~\cite{Ekstedt:2024fyq}\footnote{The relevant model file can be found in the {\tt Models/SingletStandardModel\_Z2 }.} (based on Refs.~\cite{Friedlander:2020tnq, Laurent:2022jrs}): we include the one-loop thermal functions without high-temperature expansion, and the Coleman-Weinberg potential, with a RG-scale at $125.0 \, {\rm GeV}$. We include contributions to the one-loop effective potential from the singlet, Higgs, Goldstones, $W$- and $Z$-bosons and the top quark. We take the absolute value of the masses to avoid imaginary contributions to the effective potential.
We do not consider running couplings or daisy resummation. Note that this treatment of the effective potential is expected to give rather large uncertainties \cite{Niemi:2021qvp, Gould:2021oba, Lewicki:2024xan}, but also note that a computation of the wall velocity with a consistent inclusion of higher-order corrections to the effective potential has not yet been performed, and we therefore stick here with the one-loop effective potential.

\begin{figure}[t]
    \centering
\includegraphics[width=0.49\linewidth]{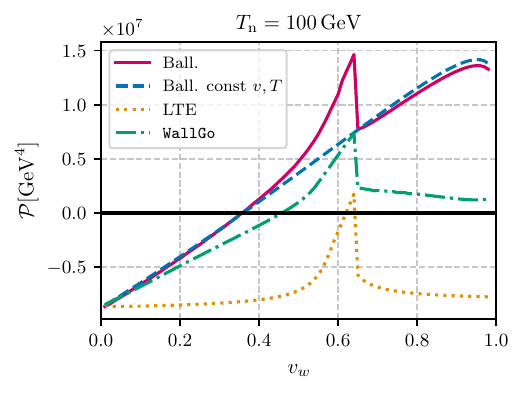}
\includegraphics[width=0.49\linewidth]{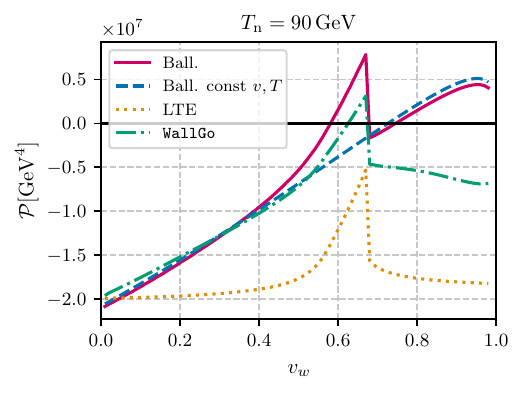}
    \caption{Pressure as a function of the wall velocity in four different approximations for the xSM.}
    \label{fig:pressurexSM}
\end{figure}

In Fig.~\ref{fig:pressurexSM}, we compare the pressure obtained in four different approaches for $T_n = 100 \, {\rm GeV}$ (left) and $T_n = 90 \, {\rm GeV}$ (right):
\begin{enumerate}
    \item The ballistic approximation, as discussed in this work (solid magenta). We include the contributions from the singlet, Higgs, Goldstone, $W$- and $Z$-bosons and the top quark. The singlet loses mass when it enters the bubble, and we therefore use the expressions of Appendix \ref{app:mass-loss} to include its contribution to the pressure.
    \item The ballistic approximation with the same contributing particles, under the approximation that $v_+ = v_- = v_w$ and $T_+ = T_- = T_n$ (dashed blue).
    \item Solution to the scalar field EoM in LTE (dotted yellow).
    \item Solution of the scalar field EoM, with friction from the top quark, without assuming the ballistic approximation (dash-dotted green).
\end{enumerate}
To obtain pressures 1, 3 and 4, we make use of the code {\tt WallGo} \cite{Ekstedt:2024fyq}. For the ballistic approximation, we only extract the hydrodynamic matching relations $(T_+,T_-,v_+,v_-)$ as a function of $v_w$ and compute the pressure using Eqs.~\eqref{eq:P-ballistic-momentum-ex}. For 3 and 4, the pressure is directly returned by {\tt WallGo}. In the computation, the bubble profile is approximated by a Tanh-ansatz, and the pressure for each wall velocity is determined by finding the wall widths and offsets that minimize the action (see Ref.~\cite{Cline:2021iff} for further details on this approach). 
Note that there is a slight ambiguity in the definition of the pressure, because the static equation of motion can not be applied for arbitrary $v_w$.
This ambiguity might explain why the different approximations do not go to exactly the same value of $P$ at $v_w \rightarrow 0$. Another possible reason for the small difference at $v_w \rightarrow 0$ is that {\tt WallGo} has been configured to return the principal part of the potential for a negative mass squared, whereas we take the absolute value in the ballistic computation.

For $T_n = 100 \, {\rm GeV}$, the four different approximations for the pressure yield $v_w = 0.36, 0.36, 0.62$, $0.46$, for ballistic, ballistic with constant $v$ and $T$, LTE and {\tt WallGo} respectively.
For $T_n = 90 \, {\rm GeV}$, we have $v_w = 0.58\ (\text{ballistic}), 0.73\ (\text{ballistic, constant $v$ and $T$}), 0.63\ (\text{\tt WallGo})$, while in LTE no static solution is found.

As anticipated, the pressure is smallest, and the wall velocity largest in the LTE approximation for both choices of nucleation temperature. We also see that the ballistic approximation with $v_w$-dependent temperature and fluid velocity gives the largest estimate of the pressure. The pressure with out-of-equilibrium top quark, with the (leading log) collision integrals falls in between the two approximations. Interestingly, the ballistic approximation of Refs.~\cite{Liu:1992tn, BarrosoMancha:2020fay} (with fixed temperature and wall velocity) reproduces $v_w$ of our updated ballistic approximation very well for $T_n= 100 \, {\rm GeV}$, but it underestimates the pressure for $T_n = 90 \, {\rm GeV}$. As a result, it overestimates the wall velocity, and finds a detonation solution rather than a hybrid.

\section{Conclusions}
\label{sec:Conc}

The LTE and ballistic approximations are two approaches for estimating bubble wall velocities in FOPTs. These approximations are typically applicable within specific regions of the parameter space and wall velocities. For instance, the LTE approximation is generally reliable for relatively low wall velocities ($\gamma_w\lsim \O(10)$) {\it and} strong interactions among plasma particles. In contrast, the ballistic approximation is suitable in scenarios with either very weak interactions or very high wall velocities (i.e., when $\gamma_w \gg 1$). In this work, however, we use these approximations to derive general bounds on friction and wall velocity. The two limits correspond to extreme collision rates: $\Gamma \rightarrow 0$ for the ballistic regime and $\Gamma \rightarrow \infty$ for LTE. By linearising the general Boltzmann equation, we have analytically demonstrated that the LTE and ballistic limits indeed establish upper and lower bounds on the wall velocity, respectively, so that $v_w \in [v_w^{\rm b}, v_w^{\rm LTE}]$.

As a key development in this work, we have demonstrated how hydrodynamics effects can be incorporated into the ballistic approximation.  To achieve this, we account for inhomogeneities in the fluid temperature and velocity while solving the Liouville equation, i.e., the collisionless Boltzmann equation. The resulting general solution depends on $\{T_\pm, v_\pm\}$, where $T_\pm$ and $v_\pm$ denote the fluid temperature and velocity in front of/behind the bubble wall, respectively. From this solution for the particle distribution functions, we compute the friction on the wall, $\P_{\rm friction}^{\rm b}(T_+, T_-, v_+, v_-)$. We have further shown that $\P_{\rm friction}^{\rm b}=\P_{\rm driving}\equiv \Delta V$ can serve as an additional matching condition, enabling us to fully determine the wall velocity. With hydrodynamics correctly integrated, we demonstrate that the ballistic pressure exhibits a non-monotonic behavior, featuring a local peak at the Jouguet velocity. This finding aligns with previous observations made using the LTE approximation or the complete solution of the Boltzmann equation~\cite{Cline:2021iff, Laurent:2022jrs, DeCurtis:2022hlx, DeCurtis:2023hil, Ai:2024shx}. 

We have conducted an extensive scan for the wall velocities obtained in the two approximations in a model-independent framework. We identify parameter regions with a small difference in the two solutions ($v^{\rm LTE}_w- v_w^{\rm b}$), where solving the full Boltzmann equation is not necessary, and those with a relatively large difference. For deflagration/hybrid motions, we have observed that in some part of parameter space, there are ballistic solutions but no LTE ones. This is reasonable as the ballistic pressure is always larger than the LTE pressure so there are situations in which the ballistic pressure can beat the driving pressure before reaching the Jouguet velocity while the LTE pressure cannot. While it has been shown in the previous work~\cite{Ai:2023see} that there is no stable detonation solution in LTE, we have shown that this is not the case for the ballistic approximation, although the parameter space allowing for such solutions is relatively narrow. For the ballistic approximation, we also provide a detailed analysis of cases where particles become very massive after entering the bubble, as occurs in models with conformal symmetry, for example. In these instances, we offer either an analytic solution for the wall velocity in the detonation case or a numerical fit for deflagration and hybrid cases. Finally, we have applied our methods to a specific model: the Standard Model coupled to a gauge singlet, and obtained the wall velocities in different limits, comparing them with the one obtained from solving the full Boltzmann equation using the package {\tt WallGo} recently released in Ref.~\cite{Ekstedt:2024fyq}. For the benchmark point we considered, we have seen that the velocity obtained by solving the full Boltzmann equation does lie between the LTE and ballistic bounds. 

Our analysis remains valid as long as hydrodynamics applies. However, we have taken Eq.~\eqref{eq:EoM-bubble} as our starting point, which accounts for the friction arising from condensate-dependent mass terms only. There may also be condensate-dependent vertices, such as $\phi \eta \chi^2$ (where $\eta$ represents the fluctuation field of the order-parameter scalar upon the bubble background), that could induce particle production processes like $\eta \rightarrow \chi \chi$. These processes and the resulting friction can be captured by self-energy modifications to Eq.~\eqref{eq:EoM-bubble}. We anticipate that these additional contributions remain subdominant at relatively low values of $\gamma_w$. For cases with $\gamma_w \rightarrow \infty$, however, these modifications would need to be taken into account. For a detailed analysis of bubble wall dynamics including self-energy corrections in the scalar EoM, see Refs.~\cite{Dashko:2024spj,Ai:forthcoming}.

\section*{Acknowledgments}
WYA is supported by the UK Engineering and Physical Sciences Research Council (EPSRC), under Research Grant No. EP/V002821/1. BL is supported by the Fonds de recherche du Québec Nature et technologies (FRQNT). JvdV is supported by the Dutch Research Council (NWO), under project number VI.Veni.212.133.

\newpage
\begin{appendix}
\renewcommand{\theequation}{\Alph{section}\arabic{equation}}

\section{Model-independent hydrodynamics with the template EoS}
\label{app:template}

To have a model-independent analysis, one usually introduces an EoS for the plasma. A good EoS is the so-called template model~\cite{Leitao:2014pda} that has been used in various studies~\cite{Giese:2020rtr,Ai:2023see,Sanchez-Garitaonandia:2023zqz}). As a generalization of the bag model, the template model allows the sound speeds in both phases to deviate from $1/\sqrt{3}$. Explicitly, it reads 
\begin{align}
\label{eq:nu_eos}
    &e_s(T)=\frac{1}{3} a_+ (\mu-1) T^\mu+\epsilon\,,\qquad p_s(T)=\frac{1}{3}a_+ T^\mu-\epsilon\,,\\
    &e_b(T)=\frac{1}{3}a_- (\nu-1)T^\nu \,,\quad\qquad\ \  p_b(T)=\frac{1}{3}a_- T^\nu\,,
\end{align}
where the constants $\mu$, $\nu$ are related to the sound speed in the symmetric and broken phases 
through
\begin{align}
    \mu=1+\frac{1}{c^2_{s}}\,,\quad \nu=1+\frac{1}{c^2_{b}}\,.
\end{align}
For $\mu=\nu=4$, it reduces to the bag EoS. 

To formulate the matching conditions in terms of model-independent parameters, we introduce the $D$ operator through~\cite{Giese:2020rtr}
\begin{align}
    \Delta p=p_s(T_+) - p_b(T_-)&=\left[p_s(T_+)-p_b(T_+)\right]+\left[p_b(T_+)-p_b(T_-)\right]\notag\\
    &\equiv Dp (T_+)+\delta p(T_+,T_-)\,,
\end{align}
and similarly for the energy density and enthalpy. Above we have defined $Dp(T)\equiv p_s(T)-p_b(T)$, the difference of a thermodynamic quantity in the two phases but at the same temperature. $\delta p$ depends on $T_+$ and $T_-$ but only involves quantities in the broken phase. A quantity that turns out to be of particular importance is
\begin{align}
    \frac{\delta p(T_+,T_-)}{\delta \rho(T_+,T_-)}=\frac{p_b(T_+)-p_b(T_-)}{e_b(T_+)-e_b(T_-)}\equiv \zeta^2(T_+,T_-)\,.
\end{align}
In the template EoS, we have $\zeta(T_+,T_-)=c_b$ which is constant.

Ref.~\cite{Giese:2020rtr, Giese:2020znk} then introduces the following quantity for the PT strength
\begin{align}
\label{eq:alpha}
    \alpha(T)=\frac{D\Bar{\theta}}{3\omega_s(T)}\,, \qquad \Bar{\theta}=\left(e-\frac{p}{c_b^2}\right)\,.
\end{align}
With the parameter $\alpha_+\equiv \alpha(T_+)$, one can write the matching conditions as
\begin{subequations}
\begin{align}
\label{eq:match0}
    &\frac{\Delta p}{3\omega_+}\left(1-\frac{v_+ v_-}{\zeta^2}\right)=v_+v_-\alpha_+\,,\\
    \label{eq:match1}
    &\frac{v_+}{v_-}=\frac{\left(\frac{v_+v_-}{\zeta^2}-1\right)+3\alpha_+}{\left(\frac{v_+v_-}{\zeta^2}-1\right)+3v_+v_-\alpha_+}\,.
\end{align}
\end{subequations}
We have now two constraint equations for five unknowns $\{v_w, v_+,v_-, T_+, T_-\}$. For detonations, $T_+$ and $v_+$ are equal to the nucleation temperature  $T_n$ and  $v_w$, respectively. For deflagrations/hybrids, $T_+$ can be related to $T_n$ through the fluid equations between the wall and the shock front, while there is an additional condition either through $v_-=v_w$, or $v_-=c_b$. One is essentially left with three unknowns. With the third constraint equation in the LTE or ballistic approximations (see Secs.~\ref{sec:LTE} and~\ref{sec:ballistic}), one then has a closed system of equations and all the unknowns can be determined, from which the wall velocity can be inferred. 

For the LTE approximation, substituting Eq.~\eqref{eq:LTE-matching-condition} into~\eqref{eq:match0} and using the EoS, one gets~\cite{Ai:2023see}  
\begin{align}
\label{eq:match2}
    3\nu\alpha_+ v_+v_- =\left[1 - 3 \alpha_{+}-\left(\frac{\gamma_+}{\gamma_-}\right)^{\nu} \Psi_+  \right] \left(1-\frac{v_+ v_-}{c_b^2}\right)\,,
\end{align}
where 
\begin{align}
\label{eq:Psi}
\Psi(T)= \omega_b(T)/\omega_s(T),
\end{align}
and $\Psi_+=\Psi(T_+)$. Now, $T_-$ has been eliminated in the two 
matching conditions~\eqref{eq:match1} (with $\zeta=c_b$) and~\eqref{eq:match2}. As $T_+$ is related to $T_n$, either directly or through the fluid equations and matching conditions at the shock front, the system of equations is closed and one can determine the wall velocity for a given $T_n$. More specifically, for detonations, $T_+=T_n$ and therefore $\alpha_+=\alpha_n\equiv \alpha(T_n)$, $\Psi_+=\Psi_n\equiv \Psi(T_n)$. For deflagrations/hybrids, one can first express $\alpha_+$ or $\Psi_+$ in terms of $\alpha_n$ or $\Psi_n$ from the template EoS
\begin{subequations}
    \begin{align}
    \alpha_+ &= \frac{\mu-\nu}{3\mu}+\frac{\omega_n}{\omega_+}\left( \alpha_n -\frac{\mu-\nu}{3\mu} \right)\,, \\
    \Psi_+ &= \Psi_n\left(\frac{\omega_+}{\omega_n}\right)^{\nu/\mu-1}\,,
\end{align}
\end{subequations}
and then compute $\omega_n/\omega_+$ by integrating the fluid equations. In conclusion, within the LTE approximation and the template model, the dynamics can be fully determined by four thermodynamic quantities: $\alpha_n$, $\Psi_n$, $c_s$ and $c_b$.

\section{Ballistic pressures for Bose-Einstein and Fermi-Dirac distributions}
\label{app:pressures-formulae}

Here we present the expressions of $\P^{t}$, $\P^{t_+}$ and $\P^{t_-}$ for Bose-Einstein and Fermi-Dirac distributions, analogous to those given in Eqs.~\eqref{eq:P-ballistic-momentum-final-ex}. We obtain
\begin{subequations}
\begin{align}
    \P^{t_+} =& \frac{T_+^4}{4\pi^2 \gamma_+}\int\limits_{x>{\sqrt{\Delta m^2}}/T_+}\d x\, x\left(x-\sqrt{x^2-\Delta m^2/T_+^2}\right)\notag\\
    &\times \left[\pm \gamma_+\left(\sqrt{x^2+m_+^2/T_+^2}-v_+ x\right) \mp \log\left(\e^{\gamma_+\left(\sqrt{x^2+m_+^2/T_+^2}-v_+ x\right) }\mp 1\right)\right]\,,\\
     \P^{t_-} =& \frac{T_-^4}{4\pi^2 \gamma_-}\int\limits_{x>0}\d x\, x\left(\sqrt{x^2+\Delta m^2/T_-^2}-x\right)\notag\\
    &\times \left[\pm \gamma_-\left(\sqrt{x^2+m_-^2/T_-^2}+v_- x\right) \mp \log\left(\e^{\gamma_-\left(\sqrt{x^2+m_-^2/T_-^2}+v_- x\right) }\mp 1\right)\right]\,,\\
    \P^r =& \frac{T_+^4}{2\pi^2\gamma_+}\int\limits_{0<x<\sqrt{\Delta m^2}/T_+}\d x\, x^2  \left[\pm \gamma_+\left(\sqrt{x^2+m_+^2/T_+^2}-v_+ x\right)\mp \log\left(\e^{\gamma_+\left(\sqrt{x^2+m_+^2/T_+^2}-v_+ x\right) }\mp 1\right)\right]\,,
\end{align}
\end{subequations}
where the upper/lower signs correspond to Bose-Einstein/Fermi-Dirac distributions.

\section{Particles that lose mass when entering the bubble}
\label{app:mass-loss}

In certain scenarios, some of the particles actually \emph{lose} mass, when entering the bubble. An example of this is the xSM, where the singlet becomes is heavier in the singlet phase than in the Higgs phase. For these particles, the distribution functions associated with transmission and reflection are given by  
\begin{itemize}
    \item[(1)] Transmission from the symmetric phase ($t_+$):
    \begin{align}
        f^{t_+}(p^z,z;\vecp_\perp)=\frac{1}{\e^{\beta_+\gamma_+\left(E-v_+\sqrt{p^{z2}+m^2(z)-m_+^2}\right)}-1}\,, \quad \left(p^z<-\sqrt{m_+^2-m^2(z)}\right)\,;
    \end{align}
    \item[(2)] Reflection ($r$): 
    \begin{align}
        f^r(p^z,z;\vecp_\perp)=&\frac{1}{\e^{\beta_-\gamma_- \left(E + v_-\sqrt{p^{z2}+m^2(z)-m_-^2}\right)}-1}\,,\notag\\&\qquad\qquad\qquad\qquad\qquad\left(-\sqrt{m_+^2-m^2(z)}<p^z<\sqrt{m_+^2-m^2(z)}\right)\,;
    \end{align} 
    \item[(3)] Transmission from inside the bubble $(t_-)$:
    \begin{align}
        f^{t_-}(p^z,z;\vecp_\perp)=\frac{1}{\e^{\beta_-\gamma_-\left(E+v_-\sqrt{p^{z2}+m^2(z)-m_-^2}\right)}-1}\,, \quad \left(p^z>\sqrt{m_+^2-m^2(z)}\right)\,.
    \end{align}
\end{itemize}
We see that $m_-$ got replaced by $m_+$ in the conditions for $p^z$ and that reflection now happens from the \emph{inside} of the bubble.

The distribution at $z \rightarrow \infty$
is 
\begin{equation}
\label{eq:f+infInvert}
f(p^z,\infty;\vecp_\perp) = \begin{cases}
    \frac{1}{\e^{\beta_+ \gamma_+ \left(E+v_+p^{z}\right) }-1}\,,\quad p^z<0\,,\quad &(\text{$t^+$-modes})\\
    \frac{1}{\e^{\beta_-\gamma_- \big(E+v_-\sqrt{p^{z2} + \Delta \overline{m}^2}\big)}-1}\,,\quad p^z>0\,,\quad &(\text{$t^-$-modes})
\end{cases}\,,
\end{equation}
where $\Delta \overline{m}^2 \equiv  m_+^2- m_-^2 = - \Delta m^2$. At $z\rightarrow -\infty$, we have 
\begin{align}
\label{eq:f-infInvert}
    f(p^z,-\infty;\vecp_\perp)=
\begin{cases}
\frac{1}{\e^{\beta_+\gamma_+\left(E-v_+ \sqrt{p^{z2} -\Delta \overline{m}^2}\right)}-1} \,,\quad p^z<-{\sqrt{\Delta \overline{m}^2}}\,,\quad &(\text{$t^+$-modes}) \\
\frac{1}{\e^{\beta_-\gamma_-\left(E + v_-|p^z| \right)}-1}\,,\quad -{\sqrt{\Delta \overline{m}^2}}<p^z<{\sqrt{\Delta \overline{m}^2}}\,,\quad &(\text{$r$-modes})\\
\frac{1}{\e^{\beta_-\gamma_-\left(E+v_-p^z\right)}-1}\,,\quad p^z> {\sqrt{\Delta \overline{m}^2}}\,, \quad &(\text{$t^-$-modes})
\end{cases}\,.
\end{align}
The different contributions to the pressure are given by
\begin{subequations}
\begin{align}
    \P^{t_+} &= \int\limits_{p^z<0}\frac{\d^3\vecp}{(2\pi)^3}\frac{(p^z)^2}{E_+}f^{t_+}(p^z,+\infty) - \int\limits_{p^z < -\sqrt{\Delta \overline{m}^2}}\frac{\d^3\vecp}{(2\pi)^3}\frac{(p^z)^2}{E_-}f^{t_+}(p^z,-\infty)\,,\\
    \P^{t_-} &= \int\limits_{p^z>0}\frac{\d^3\vecp}{(2\pi)^3}\frac{(p^z)^2}{E_+}f^{t_-}(p^z,+\infty) - \int\limits_{p^z > \sqrt{\Delta \overline{m}^2}}\frac{\d^3\vecp}{(2\pi)^3}\frac{(p^z)^2}{E_-}f^{t_-}(p^z,-\infty)\,,\\
    \P^r &= - \int\limits_{|p^z|<{\sqrt{\Delta \overline{m}^2}}}\frac{\d^3\vecp}{(2\pi)^3}\frac{(p^z)^2}{E_+} f^r(p^z,-\infty)\,.
\end{align}
\end{subequations}
These equations can be simplified by observing that $f^{t_\pm}(p^z,\pm\infty) = f^{\rm eq}(p^z,\pm\infty)$, as before, and $f^{t_\pm}(p^z,\mp\infty)=f^{\rm eq}(\mp\sqrt{p^{z2}\mp \Delta \overline{m}^2},\pm\infty)$ and $f^r(p^z,-\infty)=f^{\rm eq}(+|p^z|,-\infty)$ 
\begin{subequations}
\label{eq:P-ballistic-momentum-exInvert}
\begin{align}
    \P^{t_+} &= \int\limits_{p^z<0}\frac{\d^3\vecp}{(2\pi)^3}\frac{p^z}{E_+}\left(p^z +\sqrt{(p^z)^2+\Delta \overline{m}^2}\right)f^{\rm eq}(p^z,+\infty)\,,\\
    \P^{t_-} &= \int\limits_{p^z>\sqrt{\Delta \overline{m}^2}}\frac{\d^3\vecp}{(2\pi)^3}\frac{p^z}{E_-}\left(\sqrt{(p^z)^2-\Delta \overline{m}^2}-p^z\right)f^{\rm eq}(p^z,-\infty)\,,\\
    \label{eq:Pr-exact}
    \P^r &= - 2\int\limits_{ 0< p^z < {\sqrt{\Delta \overline{m}^2}}}\frac{\d^3\vecp}{(2\pi)^3}\frac{(p^z)^2}{E_+}f^{\rm eq}(p^z, -\infty)\,.
\end{align}
\end{subequations}
Note that all the above pressures are now negative.

\end{appendix}

\newpage
\bibliographystyle{utphys}
\bibliography{ref}{}

\providecommand{\href}[2]{#2}\begingroup\raggedright\begin{thebibliography}{100}

\bibitem{Cline:1996mga}
J.~M. Cline and P.-A. Lemieux, ``{Electroweak phase transition in two Higgs doublet models},'' \href{http://dx.doi.org/10.1103/PhysRevD.55.3873}{{\em Phys. Rev. D} {\bfseries 55} (1997) 3873--3881}, \href{http://arxiv.org/abs/hep-ph/9609240}{{\ttfamily arXiv:hep-ph/9609240}}.

\bibitem{Carena:1997ki}
M.~Carena, M.~Quiros, and C.~E.~M. Wagner, ``{Electroweak baryogenesis and Higgs and stop searches at LEP and the Tevatron},'' \href{http://dx.doi.org/10.1016/S0550-3213(98)00187-4}{{\em Nucl. Phys. B} {\bfseries 524} (1998) 3--22}, \href{http://arxiv.org/abs/hep-ph/9710401}{{\ttfamily arXiv:hep-ph/9710401}}.

\bibitem{Grojean:2004xa}
C.~Grojean, G.~Servant, and J.~D. Wells, ``{First-order electroweak phase transition in the standard model with a low cutoff},'' \href{http://dx.doi.org/10.1103/PhysRevD.71.036001}{{\em Phys. Rev. D} {\bfseries 71} (2005) 036001}, \href{http://arxiv.org/abs/hep-ph/0407019}{{\ttfamily arXiv:hep-ph/0407019}}.

\bibitem{Fromme:2006cm}
L.~Fromme, S.~J. Huber, and M.~Seniuch, ``{Baryogenesis in the two-Higgs doublet model},'' \href{http://dx.doi.org/10.1088/1126-6708/2006/11/038}{{\em JHEP} {\bfseries 11} (2006) 038}, \href{http://arxiv.org/abs/hep-ph/0605242}{{\ttfamily arXiv:hep-ph/0605242}}.

\bibitem{Grojean:2006bp}
C.~Grojean and G.~Servant, ``{Gravitational Waves from Phase Transitions at the Electroweak Scale and Beyond},'' \href{http://dx.doi.org/10.1103/PhysRevD.75.043507}{{\em Phys. Rev. D} {\bfseries 75} (2007) 043507}, \href{http://arxiv.org/abs/hep-ph/0607107}{{\ttfamily arXiv:hep-ph/0607107}}.

\bibitem{Profumo:2007wc}
S.~Profumo, M.~J. Ramsey-Musolf, and G.~Shaughnessy, ``{Singlet Higgs phenomenology and the electroweak phase transition},'' \href{http://dx.doi.org/10.1088/1126-6708/2007/08/010}{{\em JHEP} {\bfseries 08} (2007) 010}, \href{http://arxiv.org/abs/0705.2425}{{\ttfamily arXiv:0705.2425 [hep-ph]}}.

\bibitem{Barger:2007im}
V.~Barger, P.~Langacker, M.~McCaskey, M.~J. Ramsey-Musolf, and G.~Shaughnessy, ``{LHC Phenomenology of an Extended Standard Model with a Real Scalar Singlet},'' \href{http://dx.doi.org/10.1103/PhysRevD.77.035005}{{\em Phys. Rev. D} {\bfseries 77} (2008) 035005}, \href{http://arxiv.org/abs/0706.4311}{{\ttfamily arXiv:0706.4311 [hep-ph]}}.

\bibitem{Delaunay:2007wb}
C.~Delaunay, C.~Grojean, and J.~D. Wells, ``{Dynamics of Non-renormalizable Electroweak Symmetry Breaking},'' \href{http://dx.doi.org/10.1088/1126-6708/2008/04/029}{{\em JHEP} {\bfseries 04} (2008) 029}, \href{http://arxiv.org/abs/0711.2511}{{\ttfamily arXiv:0711.2511 [hep-ph]}}.

\bibitem{FileviezPerez:2008bj}
P.~Fileviez~Perez, H.~H. Patel, M.~J. Ramsey-Musolf, and K.~Wang, ``{Triplet Scalars and Dark Matter at the LHC},'' \href{http://dx.doi.org/10.1103/PhysRevD.79.055024}{{\em Phys. Rev. D} {\bfseries 79} (2009) 055024}, \href{http://arxiv.org/abs/0811.3957}{{\ttfamily arXiv:0811.3957 [hep-ph]}}.

\bibitem{Gil:2012ya}
G.~Gil, P.~Chankowski, and M.~Krawczyk, ``{Inert Dark Matter and Strong Electroweak Phase Transition},'' \href{http://dx.doi.org/10.1016/j.physletb.2012.09.052}{{\em Phys. Lett. B} {\bfseries 717} (2012) 396--402}, \href{http://arxiv.org/abs/1207.0084}{{\ttfamily arXiv:1207.0084 [hep-ph]}}.

\bibitem{Dorsch:2013wja}
G.~C. Dorsch, S.~J. Huber, and J.~M. No, ``{A strong electroweak phase transition in the 2HDM after LHC8},'' \href{http://dx.doi.org/10.1007/JHEP10(2013)029}{{\em JHEP} {\bfseries 10} (2013) 029}, \href{http://arxiv.org/abs/1305.6610}{{\ttfamily arXiv:1305.6610 [hep-ph]}}.

\bibitem{Huang:2016odd}
F.~P. Huang, Y.~Wan, D.-G. Wang, Y.-F. Cai, and X.~Zhang, ``{Hearing the echoes of electroweak baryogenesis with gravitational wave detectors},'' \href{http://dx.doi.org/10.1103/PhysRevD.94.041702}{{\em Phys. Rev. D} {\bfseries 94} no.~4, (2016) 041702}, \href{http://arxiv.org/abs/1601.01640}{{\ttfamily arXiv:1601.01640 [hep-ph]}}.

\bibitem{Jinno:2016knw}
R.~Jinno and M.~Takimoto, ``{Probing a classically conformal B-L model with gravitational waves},'' \href{http://dx.doi.org/10.1103/PhysRevD.95.015020}{{\em Phys. Rev. D} {\bfseries 95} no.~1, (2017) 015020}, \href{http://arxiv.org/abs/1604.05035}{{\ttfamily arXiv:1604.05035 [hep-ph]}}.

\bibitem{Chao:2017vrq}
W.~Chao, H.-K. Guo, and J.~Shu, ``{Gravitational Wave Signals of Electroweak Phase Transition Triggered by Dark Matter},'' \href{http://dx.doi.org/10.1088/1475-7516/2017/09/009}{{\em JCAP} {\bfseries 09} (2017) 009}, \href{http://arxiv.org/abs/1702.02698}{{\ttfamily arXiv:1702.02698 [hep-ph]}}.

\bibitem{Beniwal:2017eik}
A.~Beniwal, M.~Lewicki, J.~D. Wells, M.~White, and A.~G. Williams, ``{Gravitational wave, collider and dark matter signals from a scalar singlet electroweak baryogenesis},'' \href{http://dx.doi.org/10.1007/JHEP08(2017)108}{{\em JHEP} {\bfseries 08} (2017) 108}, \href{http://arxiv.org/abs/1702.06124}{{\ttfamily arXiv:1702.06124 [hep-ph]}}.

\bibitem{Marzola:2017jzl}
L.~Marzola, A.~Racioppi, and V.~Vaskonen, ``{Phase transition and gravitational wave phenomenology of scalar conformal extensions of the Standard Model},'' \href{http://dx.doi.org/10.1140/epjc/s10052-017-4996-1}{{\em Eur. Phys. J. C} {\bfseries 77} no.~7, (2017) 484}, \href{http://arxiv.org/abs/1704.01034}{{\ttfamily arXiv:1704.01034 [hep-ph]}}.

\bibitem{Kurup:2017dzf}
G.~Kurup and M.~Perelstein, ``{Dynamics of Electroweak Phase Transition In Singlet-Scalar Extension of the Standard Model},'' \href{http://dx.doi.org/10.1103/PhysRevD.96.015036}{{\em Phys. Rev. D} {\bfseries 96} no.~1, (2017) 015036}, \href{http://arxiv.org/abs/1704.03381}{{\ttfamily arXiv:1704.03381 [hep-ph]}}.

\bibitem{Chen:2017cyc}
Y.~Chen, M.~Huang, and Q.-S. Yan, ``{Gravitation waves from QCD and electroweak phase transitions},'' \href{http://dx.doi.org/10.1007/JHEP05(2018)178}{{\em JHEP} {\bfseries 05} (2018) 178}, \href{http://arxiv.org/abs/1712.03470}{{\ttfamily arXiv:1712.03470 [hep-ph]}}.

\bibitem{Baldes:2018emh}
I.~Baldes and C.~Garcia-Cely, ``{Strong gravitational radiation from a simple dark matter model},'' \href{http://dx.doi.org/10.1007/JHEP05(2019)190}{{\em JHEP} {\bfseries 05} (2019) 190}, \href{http://arxiv.org/abs/1809.01198}{{\ttfamily arXiv:1809.01198 [hep-ph]}}.

\bibitem{Prokopec:2018tnq}
T.~Prokopec, J.~Rezacek, and B.~\'Swie\.zewska, ``{Gravitational waves from conformal symmetry breaking},'' \href{http://dx.doi.org/10.1088/1475-7516/2019/02/009}{{\em JCAP} {\bfseries 02} (2019) 009}, \href{http://arxiv.org/abs/1809.11129}{{\ttfamily arXiv:1809.11129 [hep-ph]}}.

\bibitem{Bian:2018bxr}
L.~Bian and X.~Liu, ``{Two-step strongly first-order electroweak phase transition modified FIMP dark matter, gravitational wave signals, and the neutrino mass},'' \href{http://dx.doi.org/10.1103/PhysRevD.99.055003}{{\em Phys. Rev. D} {\bfseries 99} no.~5, (2019) 055003}, \href{http://arxiv.org/abs/1811.03279}{{\ttfamily arXiv:1811.03279 [hep-ph]}}.

\bibitem{Marzo:2018nov}
C.~Marzo, L.~Marzola, and V.~Vaskonen, ``{Phase transition and vacuum stability in the classically conformal B\textendash{}L model},'' \href{http://dx.doi.org/10.1140/epjc/s10052-019-7076-x}{{\em Eur. Phys. J. C} {\bfseries 79} no.~7, (2019) 601}, \href{http://arxiv.org/abs/1811.11169}{{\ttfamily arXiv:1811.11169 [hep-ph]}}.

\bibitem{Chala:2018opy}
M.~Chala, M.~Ramos, and M.~Spannowsky, ``{Gravitational wave and collider probes of a triplet Higgs sector with a low cutoff},'' \href{http://dx.doi.org/10.1140/epjc/s10052-019-6655-1}{{\em Eur. Phys. J. C} {\bfseries 79} no.~2, (2019) 156}, \href{http://arxiv.org/abs/1812.01901}{{\ttfamily arXiv:1812.01901 [hep-ph]}}.

\bibitem{Zhou:2018zli}
R.~Zhou, W.~Cheng, X.~Deng, L.~Bian, and Y.~Wu, ``{Electroweak phase transition and Higgs phenomenology in the Georgi-Machacek model},'' \href{http://dx.doi.org/10.1007/JHEP01(2019)216}{{\em JHEP} {\bfseries 01} (2019) 216}, \href{http://arxiv.org/abs/1812.06217}{{\ttfamily arXiv:1812.06217 [hep-ph]}}.

\bibitem{Alves:2018jsw}
A.~Alves, T.~Ghosh, H.-K. Guo, K.~Sinha, and D.~Vagie, ``{Collider and Gravitational Wave Complementarity in Exploring the Singlet Extension of the Standard Model},'' \href{http://dx.doi.org/10.1007/JHEP04(2019)052}{{\em JHEP} {\bfseries 04} (2019) 052}, \href{http://arxiv.org/abs/1812.09333}{{\ttfamily arXiv:1812.09333 [hep-ph]}}.

\bibitem{Azatov:2019png}
A.~Azatov, D.~Barducci, and F.~Sgarlata, ``{Gravitational traces of broken gauge symmetries},'' \href{http://dx.doi.org/10.1088/1475-7516/2020/07/027}{{\em JCAP} {\bfseries 07} (2020) 027}, \href{http://arxiv.org/abs/1910.01124}{{\ttfamily arXiv:1910.01124 [hep-ph]}}.

\bibitem{DelleRose:2019pgi}
L.~Delle~Rose, G.~Panico, M.~Redi, and A.~Tesi, ``{Gravitational Waves from Supercool Axions},'' \href{http://dx.doi.org/10.1007/JHEP04(2020)025}{{\em JHEP} {\bfseries 04} (2020) 025}, \href{http://arxiv.org/abs/1912.06139}{{\ttfamily arXiv:1912.06139 [hep-ph]}}.

\bibitem{VonHarling:2019rgb}
B.~Von~Harling, A.~Pomarol, O.~Pujol\`as, and F.~Rompineve, ``{Peccei-Quinn Phase Transition at LIGO},'' \href{http://dx.doi.org/10.1007/JHEP04(2020)195}{{\em JHEP} {\bfseries 04} (2020) 195}, \href{http://arxiv.org/abs/1912.07587}{{\ttfamily arXiv:1912.07587 [hep-ph]}}.

\bibitem{Halverson:2020xpg}
J.~Halverson, C.~Long, A.~Maiti, B.~Nelson, and G.~Salinas, ``{Gravitational waves from dark Yang-Mills sectors},'' \href{http://dx.doi.org/10.1007/JHEP05(2021)154}{{\em JHEP} {\bfseries 05} (2021) 154}, \href{http://arxiv.org/abs/2012.04071}{{\ttfamily arXiv:2012.04071 [hep-ph]}}.

\bibitem{Ghosh:2020ipy}
T.~Ghosh, H.-K. Guo, T.~Han, and H.~Liu, ``{Electroweak phase transition with an SU(2) dark sector},'' \href{http://dx.doi.org/10.1007/JHEP07(2021)045}{{\em JHEP} {\bfseries 07} (2021) 045}, \href{http://arxiv.org/abs/2012.09758}{{\ttfamily arXiv:2012.09758 [hep-ph]}}.

\bibitem{Huang:2020crf}
W.-C. Huang, M.~Reichert, F.~Sannino, and Z.-W. Wang, ``{Testing the dark SU(N) Yang-Mills theory confined landscape: From the lattice to gravitational waves},'' \href{http://dx.doi.org/10.1103/PhysRevD.104.035005}{{\em Phys. Rev. D} {\bfseries 104} no.~3, (2021) 035005}, \href{http://arxiv.org/abs/2012.11614}{{\ttfamily arXiv:2012.11614 [hep-ph]}}.

\bibitem{DiBari:2021dri}
P.~Di~Bari, D.~Marfatia, and Y.-L. Zhou, ``{Gravitational waves from first-order phase transitions in Majoron models of neutrino mass},'' \href{http://dx.doi.org/10.1007/JHEP10(2021)193}{{\em JHEP} {\bfseries 10} (2021) 193}, \href{http://arxiv.org/abs/2106.00025}{{\ttfamily arXiv:2106.00025 [hep-ph]}}.

\bibitem{Kierkla:2022odc}
M.~Kierkla, A.~Karam, and B.~Swiezewska, ``{Conformal model for gravitational waves and dark matter: a status update},'' \href{http://dx.doi.org/10.1007/JHEP03(2023)007}{{\em JHEP} {\bfseries 03} (2023) 007}, \href{http://arxiv.org/abs/2210.07075}{{\ttfamily arXiv:2210.07075 [astro-ph.CO]}}.

\bibitem{Morgante:2022zvc}
E.~Morgante, N.~Ramberg, and P.~Schwaller, ``{Gravitational waves from dark SU(3) Yang-Mills theory},'' \href{http://dx.doi.org/10.1103/PhysRevD.107.036010}{{\em Phys. Rev. D} {\bfseries 107} no.~3, (2023) 036010}, \href{http://arxiv.org/abs/2210.11821}{{\ttfamily arXiv:2210.11821 [hep-ph]}}.

\bibitem{Fujikura:2023fbi}
K.~Fujikura, Y.~Nakai, R.~Sato, and Y.~Wang, ``{Cosmological phase transitions in composite Higgs models},'' \href{http://dx.doi.org/10.1007/JHEP09(2023)053}{{\em JHEP} {\bfseries 09} (2023) 053}, \href{http://arxiv.org/abs/2306.01305}{{\ttfamily arXiv:2306.01305 [hep-ph]}}.

\bibitem{Frandsen:2023vhu}
M.~T. Frandsen, M.~Heikinheimo, M.~Rosenlyst, M.~E. Thing, and K.~Tuominen, ``{Gravitational waves from SU(N)/SP(N) composite Higgs models},'' \href{http://dx.doi.org/10.1007/JHEP09(2023)022}{{\em JHEP} {\bfseries 09} (2023) 022}, \href{http://arxiv.org/abs/2302.09104}{{\ttfamily arXiv:2302.09104 [hep-ph]}}.

\bibitem{Pasechnik:2023hwv}
R.~Pasechnik, M.~Reichert, F.~Sannino, and Z.-W. Wang, ``{Gravitational waves from composite dark sectors},'' \href{http://dx.doi.org/10.1007/JHEP02(2024)159}{{\em JHEP} {\bfseries 02} (2024) 159}, \href{http://arxiv.org/abs/2309.16755}{{\ttfamily arXiv:2309.16755 [hep-ph]}}.

\bibitem{Feng:2024pab}
W.-Z. Feng, J.~Li, and P.~Nath, ``{Cosmologically consistent analysis of gravitational waves from hidden sectors},'' \href{http://dx.doi.org/10.1103/PhysRevD.110.015020}{{\em Phys. Rev. D} {\bfseries 110} no.~1, (2024) 015020}, \href{http://arxiv.org/abs/2403.09558}{{\ttfamily arXiv:2403.09558 [hep-ph]}}.

\bibitem{Gao:2024pqm}
F.~Gao, S.~Sun, and G.~White, ``{A first-order deconfinement phase transition in the early universe and gravitational waves},'' \href{http://arxiv.org/abs/2405.00490}{{\ttfamily arXiv:2405.00490 [hep-ph]}}.

\bibitem{Gao:2024fhm}
F.~Gao, J.~Harz, C.~Hati, Y.~Lu, I.~M. Oldengott, and G.~White, ``{Baryogenesis and first-order QCD transition with gravitational waves from a large lepton asymmetry},'' \href{http://arxiv.org/abs/2407.17549}{{\ttfamily arXiv:2407.17549 [hep-ph]}}.

\bibitem{Witten:1984rs}
E.~Witten, ``{Cosmic Separation of Phases},'' \href{http://dx.doi.org/10.1103/PhysRevD.30.272}{{\em Phys. Rev. D} {\bfseries 30} (1984) 272--285}.

\bibitem{Hogan:1986dsh}
C.~J. Hogan, ``{Gravitational radiation from cosmological phase transitions},'' \href{http://dx.doi.org/10.1093/mnras/218.4.629}{{\em Mon. Not. Roy. Astron. Soc.} {\bfseries 218} no.~4, (1986) 629--636}.

\bibitem{Kosowsky:1992vn}
A.~Kosowsky and M.~S. Turner, ``{Gravitational radiation from colliding vacuum bubbles: envelope approximation to many bubble collisions},'' \href{http://dx.doi.org/10.1103/PhysRevD.47.4372}{{\em Phys. Rev. D} {\bfseries 47} (1993) 4372--4391}, \href{http://arxiv.org/abs/astro-ph/9211004}{{\ttfamily arXiv:astro-ph/9211004}}.

\bibitem{Kosowsky:1992rz}
A.~Kosowsky, M.~S. Turner, and R.~Watkins, ``{Gravitational waves from first order cosmological phase transitions},'' \href{http://dx.doi.org/10.1103/PhysRevLett.69.2026}{{\em Phys. Rev. Lett.} {\bfseries 69} (1992) 2026--2029}.

\bibitem{Kamionkowski:1993fg}
M.~Kamionkowski, A.~Kosowsky, and M.~S. Turner, ``{Gravitational radiation from first order phase transitions},'' \href{http://dx.doi.org/10.1103/PhysRevD.49.2837}{{\em Phys. Rev. D} {\bfseries 49} (1994) 2837--2851}, \href{http://arxiv.org/abs/astro-ph/9310044}{{\ttfamily arXiv:astro-ph/9310044}}.

\bibitem{Kuzmin:1985mm}
V.~A. Kuzmin, V.~A. Rubakov, and M.~E. Shaposhnikov, ``{On the Anomalous Electroweak Baryon Number Nonconservation in the Early Universe},'' \href{http://dx.doi.org/10.1016/0370-2693(85)91028-7}{{\em Phys. Lett. B} {\bfseries 155} (1985) 36}.

\bibitem{Morrissey:2012db}
D.~E. Morrissey and M.~J. Ramsey-Musolf, ``{Electroweak baryogenesis},'' \href{http://dx.doi.org/10.1088/1367-2630/14/12/125003}{{\em New J. Phys.} {\bfseries 14} (2012) 125003}, \href{http://arxiv.org/abs/1206.2942}{{\ttfamily arXiv:1206.2942 [hep-ph]}}.

\bibitem{Garbrecht:2018mrp}
B.~Garbrecht, ``{Why is there more matter than antimatter? Calculational methods for leptogenesis and electroweak baryogenesis},'' \href{http://dx.doi.org/10.1016/j.ppnp.2019.103727}{{\em Prog. Part. Nucl. Phys.} {\bfseries 110} (2020) 103727}, \href{http://arxiv.org/abs/1812.02651}{{\ttfamily arXiv:1812.02651 [hep-ph]}}.

\bibitem{Cline:2020jre}
J.~M. Cline and K.~Kainulainen, ``{Electroweak baryogenesis at high bubble wall velocities},'' \href{http://dx.doi.org/10.1103/PhysRevD.101.063525}{{\em Phys. Rev. D} {\bfseries 101} no.~6, (2020) 063525}, \href{http://arxiv.org/abs/2001.00568}{{\ttfamily arXiv:2001.00568 [hep-ph]}}.

\bibitem{Azatov:2021irb}
A.~Azatov, M.~Vanvlasselaer, and W.~Yin, ``{Baryogenesis via relativistic bubble walls},'' \href{http://dx.doi.org/10.1007/JHEP10(2021)043}{{\em JHEP} {\bfseries 10} (2021) 043}, \href{http://arxiv.org/abs/2106.14913}{{\ttfamily arXiv:2106.14913 [hep-ph]}}.

\bibitem{Baldes:2021vyz}
I.~Baldes, S.~Blasi, A.~Mariotti, A.~Sevrin, and K.~Turbang, ``{Baryogenesis via relativistic bubble expansion},'' \href{http://dx.doi.org/10.1103/PhysRevD.104.115029}{{\em Phys. Rev. D} {\bfseries 104} no.~11, (2021) 115029}, \href{http://arxiv.org/abs/2106.15602}{{\ttfamily arXiv:2106.15602 [hep-ph]}}.

\bibitem{Huang:2022vkf}
P.~Huang and K.-P. Xie, ``{Leptogenesis triggered by a first-order phase transition},'' \href{http://dx.doi.org/10.1007/JHEP09(2022)052}{{\em JHEP} {\bfseries 09} (2022) 052}, \href{http://arxiv.org/abs/2206.04691}{{\ttfamily arXiv:2206.04691 [hep-ph]}}.

\bibitem{Chun:2023ezg}
E.~J. Chun, T.~P. Dutka, T.~H. Jung, X.~Nagels, and M.~Vanvlasselaer, ``{Bubble-assisted leptogenesis},'' \href{http://dx.doi.org/10.1007/JHEP09(2023)164}{{\em JHEP} {\bfseries 09} (2023) 164}, \href{http://arxiv.org/abs/2305.10759}{{\ttfamily arXiv:2305.10759 [hep-ph]}}.

\bibitem{Cataldi:2024pgt}
M.~Cataldi and B.~Shakya, ``{Leptogenesis via Bubble Collisions},'' \href{http://arxiv.org/abs/2407.16747}{{\ttfamily arXiv:2407.16747 [hep-ph]}}.

\bibitem{Falkowski:2012fb}
A.~Falkowski and J.~M. No, ``{Non-thermal Dark Matter Production from the Electroweak Phase Transition: Multi-TeV WIMPs and 'Baby-Zillas'},'' \href{http://dx.doi.org/10.1007/JHEP02(2013)034}{{\em JHEP} {\bfseries 02} (2013) 034}, \href{http://arxiv.org/abs/1211.5615}{{\ttfamily arXiv:1211.5615 [hep-ph]}}.

\bibitem{Baker:2019ndr}
M.~J. Baker, J.~Kopp, and A.~J. Long, ``{Filtered Dark Matter at a First Order Phase Transition},'' \href{http://dx.doi.org/10.1103/PhysRevLett.125.151102}{{\em Phys. Rev. Lett.} {\bfseries 125} no.~15, (2020) 151102}, \href{http://arxiv.org/abs/1912.02830}{{\ttfamily arXiv:1912.02830 [hep-ph]}}.

\bibitem{Chway:2019kft}
D.~Chway, T.~H. Jung, and C.~S. Shin, ``{Dark matter filtering-out effect during a first-order phase transition},'' \href{http://dx.doi.org/10.1103/PhysRevD.101.095019}{{\em Phys. Rev. D} {\bfseries 101} no.~9, (2020) 095019}, \href{http://arxiv.org/abs/1912.04238}{{\ttfamily arXiv:1912.04238 [hep-ph]}}.

\bibitem{Chao:2020adk}
W.~Chao, X.-F. Li, and L.~Wang, ``{Filtered pseudo-scalar dark matter and gravitational waves from first order phase transition},'' \href{http://dx.doi.org/10.1088/1475-7516/2021/06/038}{{\em JCAP} {\bfseries 06} (2021) 038}, \href{http://arxiv.org/abs/2012.15113}{{\ttfamily arXiv:2012.15113 [hep-ph]}}.

\bibitem{Azatov:2021ifm}
A.~Azatov, M.~Vanvlasselaer, and W.~Yin, ``{Dark Matter production from relativistic bubble walls},'' \href{http://dx.doi.org/10.1007/JHEP03(2021)288}{{\em JHEP} {\bfseries 03} (2021) 288}, \href{http://arxiv.org/abs/2101.05721}{{\ttfamily arXiv:2101.05721 [hep-ph]}}.

\bibitem{Azatov:2022tii}
A.~Azatov, G.~Barni, S.~Chakraborty, M.~Vanvlasselaer, and W.~Yin, ``{Ultra-relativistic bubbles from the simplest Higgs portal and their cosmological consequences},'' \href{http://dx.doi.org/10.1007/JHEP10(2022)017}{{\em JHEP} {\bfseries 10} (2022) 017}, \href{http://arxiv.org/abs/2207.02230}{{\ttfamily arXiv:2207.02230 [hep-ph]}}.

\bibitem{Baldes:2022oev}
I.~Baldes, Y.~Gouttenoire, and F.~Sala, ``{Hot and heavy dark matter from a weak scale phase transition},'' \href{http://dx.doi.org/10.21468/SciPostPhys.14.3.033}{{\em SciPost Phys.} {\bfseries 14} no.~3, (2023) 033}, \href{http://arxiv.org/abs/2207.05096}{{\ttfamily arXiv:2207.05096 [hep-ph]}}.

\bibitem{Giudice:2024tcp}
G.~F. Giudice, H.~M. Lee, A.~Pomarol, and B.~Shakya, ``{Nonthermal Heavy Dark Matter from a First-Order Phase Transition},'' \href{http://arxiv.org/abs/2403.03252}{{\ttfamily arXiv:2403.03252 [hep-ph]}}.

\bibitem{Azatov:2024crd}
A.~Azatov, X.~Nagels, M.~Vanvlasselaer, and W.~Yin, ``{Populating secluded dark sector with ultra-relativistic bubbles},'' \href{http://arxiv.org/abs/2406.12554}{{\ttfamily arXiv:2406.12554 [hep-ph]}}.

\bibitem{Ai:2024ikj}
W.-Y. Ai, M.~Fairbairn, K.~Mimasu, and T.~You, ``{Non-thermal production of heavy vector dark matter from relativistic bubble walls},'' \href{http://arxiv.org/abs/2406.20051}{{\ttfamily arXiv:2406.20051 [hep-ph]}}.

\bibitem{BarrosoMancha:2020fay}
M.~Barroso~Mancha, T.~Prokopec, and B.~Swiezewska, ``{Field-theoretic derivation of bubble-wall force},'' \href{http://dx.doi.org/10.1007/JHEP01(2021)070}{{\em JHEP} {\bfseries 01} (2021) 070}, \href{http://arxiv.org/abs/2005.10875}{{\ttfamily arXiv:2005.10875 [hep-th]}}.

\bibitem{Baldes:2020kam}
I.~Baldes, Y.~Gouttenoire, and F.~Sala, ``{String Fragmentation in Supercooled Confinement and Implications for Dark Matter},'' \href{http://dx.doi.org/10.1007/JHEP04(2021)278}{{\em JHEP} {\bfseries 04} (2021) 278}, \href{http://arxiv.org/abs/2007.08440}{{\ttfamily arXiv:2007.08440 [hep-ph]}}.

\bibitem{Friedlander:2020tnq}
A.~Friedlander, I.~Banta, J.~M. Cline, and D.~Tucker-Smith, ``{Wall speed and shape in singlet-assisted strong electroweak phase transitions},'' \href{http://dx.doi.org/10.1103/PhysRevD.103.055020}{{\em Phys. Rev. D} {\bfseries 103} no.~5, (2021) 055020}, \href{http://arxiv.org/abs/2009.14295}{{\ttfamily arXiv:2009.14295 [hep-ph]}}.

\bibitem{Balaji:2020yrx}
S.~Balaji, M.~Spannowsky, and C.~Tamarit, ``{Cosmological bubble friction in local equilibrium},'' \href{http://dx.doi.org/10.1088/1475-7516/2021/03/051}{{\em JCAP} {\bfseries 03} (2021) 051}, \href{http://arxiv.org/abs/2010.08013}{{\ttfamily arXiv:2010.08013 [hep-ph]}}.

\bibitem{Cline:2021iff}
J.~M. Cline, A.~Friedlander, D.-M. He, K.~Kainulainen, B.~Laurent, and D.~Tucker-Smith, ``{Baryogenesis and gravity waves from a UV-completed electroweak phase transition},'' \href{http://dx.doi.org/10.1103/PhysRevD.103.123529}{{\em Phys. Rev. D} {\bfseries 103} no.~12, (2021) 123529}, \href{http://arxiv.org/abs/2102.12490}{{\ttfamily arXiv:2102.12490 [hep-ph]}}.

\bibitem{Bea:2021zsu}
Y.~Bea, J.~Casalderrey-Solana, T.~Giannakopoulos, D.~Mateos, M.~Sanchez-Garitaonandia, and M.~Zilh\~ao, ``{Bubble wall velocity from holography},'' \href{http://dx.doi.org/10.1103/PhysRevD.104.L121903}{{\em Phys. Rev. D} {\bfseries 104} no.~12, (2021) L121903}, \href{http://arxiv.org/abs/2104.05708}{{\ttfamily arXiv:2104.05708 [hep-th]}}.

\bibitem{Bigazzi:2021ucw}
F.~Bigazzi, A.~Caddeo, T.~Canneti, and A.~L. Cotrone, ``{Bubble wall velocity at strong coupling},'' \href{http://dx.doi.org/10.1007/JHEP08(2021)090}{{\em JHEP} {\bfseries 08} (2021) 090}, \href{http://arxiv.org/abs/2104.12817}{{\ttfamily arXiv:2104.12817 [hep-ph]}}.

\bibitem{Ai:2021kak}
W.-Y. Ai, B.~Garbrecht, and C.~Tamarit, ``{Bubble wall velocities in local equilibrium},'' \href{http://dx.doi.org/10.1088/1475-7516/2022/03/015}{{\em JCAP} {\bfseries 03} no.~03, (2022) 015}, \href{http://arxiv.org/abs/2109.13710}{{\ttfamily arXiv:2109.13710 [hep-ph]}}.

\bibitem{Lewicki:2021pgr}
M.~Lewicki, M.~Merchand, and M.~Zych, ``{Electroweak bubble wall expansion: gravitational waves and baryogenesis in Standard Model-like thermal plasma},'' \href{http://dx.doi.org/10.1007/JHEP02(2022)017}{{\em JHEP} {\bfseries 02} (2022) 017}, \href{http://arxiv.org/abs/2111.02393}{{\ttfamily arXiv:2111.02393 [astro-ph.CO]}}.

\bibitem{Dorsch:2021nje}
G.~C. Dorsch, S.~J. Huber, and T.~Konstandin, ``{A sonic boom in bubble wall friction},'' \href{http://dx.doi.org/10.1088/1475-7516/2022/04/010}{{\em JCAP} {\bfseries 04} no.~04, (2022) 010}, \href{http://arxiv.org/abs/2112.12548}{{\ttfamily arXiv:2112.12548 [hep-ph]}}.

\bibitem{DeCurtis:2022hlx}
S.~De~Curtis, L.~D. Rose, A.~Guiggiani, A.~G. Muyor, and G.~Panico, ``{Bubble wall dynamics at the electroweak phase transition},'' \href{http://dx.doi.org/10.1007/JHEP03(2022)163}{{\em JHEP} {\bfseries 03} (2022) 163}, \href{http://arxiv.org/abs/2201.08220}{{\ttfamily arXiv:2201.08220 [hep-ph]}}.

\bibitem{Laurent:2022jrs}
B.~Laurent and J.~M. Cline, ``{First principles determination of bubble wall velocity},'' \href{http://dx.doi.org/10.1103/PhysRevD.106.023501}{{\em Phys. Rev. D} {\bfseries 106} no.~2, (2022) 023501}, \href{http://arxiv.org/abs/2204.13120}{{\ttfamily arXiv:2204.13120 [hep-ph]}}.

\bibitem{Wang:2022txy}
S.-J. Wang and Z.-Y. Yuwen, ``{Hydrodynamic backreaction force of cosmological bubble expansion},'' \href{http://dx.doi.org/10.1103/PhysRevD.107.023501}{{\em Phys. Rev. D} {\bfseries 107} no.~2, (2023) 023501}, \href{http://arxiv.org/abs/2205.02492}{{\ttfamily arXiv:2205.02492 [hep-ph]}}.

\bibitem{Lewicki:2022nba}
M.~Lewicki, V.~Vaskonen, and H.~Veerm\"ae, ``{Bubble dynamics in fluids with N-body simulations},'' \href{http://dx.doi.org/10.1103/PhysRevD.106.103501}{{\em Phys. Rev. D} {\bfseries 106} no.~10, (2022) 103501}, \href{http://arxiv.org/abs/2205.05667}{{\ttfamily arXiv:2205.05667 [astro-ph.CO]}}.

\bibitem{Janik:2022wsx}
R.~A. Janik, M.~Jarvinen, H.~Soltanpanahi, and J.~Sonnenschein, ``{Perfect Fluid Hydrodynamic Picture of Domain Wall Velocities at Strong Coupling},'' \href{http://dx.doi.org/10.1103/PhysRevLett.129.081601}{{\em Phys. Rev. Lett.} {\bfseries 129} no.~8, (2022) 081601}, \href{http://arxiv.org/abs/2205.06274}{{\ttfamily arXiv:2205.06274 [hep-th]}}.

\bibitem{Li:2023xto}
L.~Li, S.-J. Wang, and Z.-Y. Yuwen, ``{Bubble expansion at strong coupling},'' \href{http://dx.doi.org/10.1103/PhysRevD.108.096033}{{\em Phys. Rev. D} {\bfseries 108} no.~9, (2023) 096033}, \href{http://arxiv.org/abs/2302.10042}{{\ttfamily arXiv:2302.10042 [hep-th]}}.

\bibitem{Ai:2023see}
W.-Y. Ai, B.~Laurent, and J.~van~de Vis, ``{Model-independent bubble wall velocities in local thermal equilibrium},'' \href{http://dx.doi.org/10.1088/1475-7516/2023/07/002}{{\em JCAP} {\bfseries 07} (2023) 002}, \href{http://arxiv.org/abs/2303.10171}{{\ttfamily arXiv:2303.10171 [astro-ph.CO]}}.

\bibitem{Krajewski:2023clt}
T.~Krajewski, M.~Lewicki, and M.~Zych, ``{Hydrodynamical constraints on the bubble wall velocity},'' \href{http://dx.doi.org/10.1103/PhysRevD.108.103523}{{\em Phys. Rev. D} {\bfseries 108} no.~10, (2023) 103523}, \href{http://arxiv.org/abs/2303.18216}{{\ttfamily arXiv:2303.18216 [astro-ph.CO]}}.

\bibitem{Giombi:2023jqq}
L.~Giombi and M.~Hindmarsh, ``{General relativistic bubble growth in cosmological phase transitions},'' \href{http://dx.doi.org/10.1088/1475-7516/2024/03/059}{{\em JCAP} {\bfseries 03} (2024) 059}, \href{http://arxiv.org/abs/2307.12080}{{\ttfamily arXiv:2307.12080 [astro-ph.CO]}}.

\bibitem{Wang:2023kux}
J.-C. Wang, Z.-Y. Yuwen, Y.-S. Hao, and S.-J. Wang, ``{General backreaction force of cosmological bubble expansion},'' \href{http://dx.doi.org/10.1103/PhysRevD.110.016031}{{\em Phys. Rev. D} {\bfseries 110} no.~1, (2024) 016031}, \href{http://arxiv.org/abs/2310.07691}{{\ttfamily arXiv:2310.07691 [hep-ph]}}.

\bibitem{Gouttenoire:2023roe}
Y.~Gouttenoire, E.~Kuflik, and D.~Liu, ``{Heavy baryon dark matter from SU(N) confinement: Bubble wall velocity and boundary effects},'' \href{http://dx.doi.org/10.1103/PhysRevD.109.035002}{{\em Phys. Rev. D} {\bfseries 109} no.~3, (2024) 035002}, \href{http://arxiv.org/abs/2311.00029}{{\ttfamily arXiv:2311.00029 [hep-ph]}}.

\bibitem{Dorsch:2023tss}
G.~C. Dorsch and D.~A. Pinto, ``{Bubble wall velocities with an extended fluid Ansatz},'' \href{http://dx.doi.org/10.1088/1475-7516/2024/04/027}{{\em JCAP} {\bfseries 04} (2024) 027}, \href{http://arxiv.org/abs/2312.02354}{{\ttfamily arXiv:2312.02354 [hep-ph]}}.

\bibitem{Sanchez-Garitaonandia:2023zqz}
M.~Sanchez-Garitaonandia and J.~van~de Vis, ``{Prediction of the bubble wall velocity for a large jump in degrees of freedom},'' \href{http://dx.doi.org/10.1103/PhysRevD.110.023509}{{\em Phys. Rev. D} {\bfseries 110} no.~2, (2024) 023509}, \href{http://arxiv.org/abs/2312.09964}{{\ttfamily arXiv:2312.09964 [hep-ph]}}.

\bibitem{DeCurtis:2024hvh}
S.~De~Curtis, L.~Delle~Rose, A.~Guiggiani, A.~Gil~Muyor, and G.~Panico, ``{Non-linearities in cosmological bubble wall dynamics},'' \href{http://dx.doi.org/10.1007/JHEP05(2024)009}{{\em JHEP} {\bfseries 05} (2024) 009}, \href{http://arxiv.org/abs/2401.13522}{{\ttfamily arXiv:2401.13522 [hep-ph]}}.

\bibitem{Kang:2024xqk}
Z.~Kang and J.~Zhu, ``{Confinement Bubble Wall Velocity via Quasiparticle Determination},'' \href{http://arxiv.org/abs/2401.03849}{{\ttfamily arXiv:2401.03849 [hep-ph]}}.

\bibitem{Wang:2024wcs}
D.-W. Wang, Q.-S. Yan, and M.~Huang, ``{Bubble wall velocity and gravitational wave in the minimal left-right symmetric model},'' \href{http://dx.doi.org/10.1103/PhysRevD.110.076011}{{\em Phys. Rev. D} {\bfseries 110} no.~7, (2024) 076011}, \href{http://arxiv.org/abs/2405.01949}{{\ttfamily arXiv:2405.01949 [gr-qc]}}.

\bibitem{Evans:2024ilx}
N.~Evans and W.~Fan, ``{Designer bubble walls in a holographic Weyl semi-metal with magnetic field},'' \href{http://arxiv.org/abs/2408.10835}{{\ttfamily arXiv:2408.10835 [hep-th]}}.

\bibitem{Yuwen:2024hme}
Z.-Y. Yuwen, J.-C. Wang, and S.-J. Wang, ``{Bubble wall velocity from number density current in (non)equilibrium},'' \href{http://arxiv.org/abs/2409.20045}{{\ttfamily arXiv:2409.20045 [hep-ph]}}.

\bibitem{Ekstedt:2024fyq}
A.~Ekstedt, O.~Gould, J.~Hirvonen, B.~Laurent, L.~Niemi, P.~Schicho, and J.~van~de Vis, ``{How fast does the WallGo? A package for computing wall velocities in first-order phase transitions},'' \href{http://arxiv.org/abs/2411.04970}{{\ttfamily arXiv:2411.04970 [hep-ph]}}.

\bibitem{Moore:1995si}
G.~D. Moore and T.~Prokopec, ``{How fast can the wall move? A Study of the electroweak phase transition dynamics},'' \href{http://dx.doi.org/10.1103/PhysRevD.52.7182}{{\em Phys. Rev. D} {\bfseries 52} (1995) 7182--7204}, \href{http://arxiv.org/abs/hep-ph/9506475}{{\ttfamily arXiv:hep-ph/9506475}}.

\bibitem{Moore:1995ua}
G.~D. Moore and T.~Prokopec, ``{Bubble wall velocity in a first order electroweak phase transition},'' \href{http://dx.doi.org/10.1103/PhysRevLett.75.777}{{\em Phys. Rev. Lett.} {\bfseries 75} (1995) 777--780}, \href{http://arxiv.org/abs/hep-ph/9503296}{{\ttfamily arXiv:hep-ph/9503296}}.

\bibitem{Liu:1992tn}
B.-H. Liu, L.~D. McLerran, and N.~Turok, ``{Bubble nucleation and growth at a baryon number producing electroweak phase transition},'' \href{http://dx.doi.org/10.1103/PhysRevD.46.2668}{{\em Phys. Rev. D} {\bfseries 46} (1992) 2668--2688}.

\bibitem{Dorsch:2018pat}
G.~C. Dorsch, S.~J. Huber, and T.~Konstandin, ``{Bubble wall velocities in the Standard Model and beyond},'' \href{http://dx.doi.org/10.1088/1475-7516/2018/12/034}{{\em JCAP} {\bfseries 12} (2018) 034}, \href{http://arxiv.org/abs/1809.04907}{{\ttfamily arXiv:1809.04907 [hep-ph]}}.

\bibitem{Wang:2020zlf}
X.~Wang, F.~P. Huang, and X.~Zhang, ``{Bubble wall velocity beyond leading-log approximation in electroweak phase transition},'' \href{http://arxiv.org/abs/2011.12903}{{\ttfamily arXiv:2011.12903 [hep-ph]}}.

\bibitem{Laurent:2020gpg}
B.~Laurent and J.~M. Cline, ``{Fluid equations for fast-moving electroweak bubble walls},'' \href{http://dx.doi.org/10.1103/PhysRevD.102.063516}{{\em Phys. Rev. D} {\bfseries 102} no.~6, (2020) 063516}, \href{http://arxiv.org/abs/2007.10935}{{\ttfamily arXiv:2007.10935 [hep-ph]}}.

\bibitem{Jiang:2022btc}
S.~Jiang, F.~P. Huang, and X.~Wang, ``{Bubble wall velocity during electroweak phase transition in the inert doublet model},'' \href{http://dx.doi.org/10.1103/PhysRevD.107.095005}{{\em Phys. Rev. D} {\bfseries 107} no.~9, (2023) 095005}, \href{http://arxiv.org/abs/2211.13142}{{\ttfamily arXiv:2211.13142 [hep-ph]}}.

\bibitem{Ignatius:1993qn}
J.~Ignatius, K.~Kajantie, H.~Kurki-Suonio, and M.~Laine, ``{The growth of bubbles in cosmological phase transitions},'' \href{http://dx.doi.org/10.1103/PhysRevD.49.3854}{{\em Phys. Rev. D} {\bfseries 49} (1994) 3854--3868}, \href{http://arxiv.org/abs/astro-ph/9309059}{{\ttfamily arXiv:astro-ph/9309059}}.

\bibitem{Heckler:1994uu}
A.~F. Heckler, ``{The Effects of electroweak phase transition dynamics on baryogenesis and primordial nucleosynthesis},'' \href{http://dx.doi.org/10.1103/PhysRevD.51.405}{{\em Phys. Rev. D} {\bfseries 51} (1995) 405--428}, \href{http://arxiv.org/abs/astro-ph/9407064}{{\ttfamily arXiv:astro-ph/9407064}}.

\bibitem{Kurki-Suonio:1996gkq}
H.~Kurki-Suonio and M.~Laine, ``{Real time history of the cosmological electroweak phase transition},'' \href{http://dx.doi.org/10.1103/PhysRevLett.77.3951}{{\em Phys. Rev. Lett.} {\bfseries 77} (1996) 3951--3954}, \href{http://arxiv.org/abs/hep-ph/9607382}{{\ttfamily arXiv:hep-ph/9607382}}.

\bibitem{Espinosa:2010hh}
J.~R. Espinosa, T.~Konstandin, J.~M. No, and G.~Servant, ``{Energy Budget of Cosmological First-order Phase Transitions},'' \href{http://dx.doi.org/10.1088/1475-7516/2010/06/028}{{\em JCAP} {\bfseries 06} (2010) 028}, \href{http://arxiv.org/abs/1004.4187}{{\ttfamily arXiv:1004.4187 [hep-ph]}}.

\bibitem{Huber:2011aa}
S.~J. Huber and M.~Sopena, ``{The bubble wall velocity in the minimal supersymmetric light stop scenario},'' \href{http://dx.doi.org/10.1103/PhysRevD.85.103507}{{\em Phys. Rev. D} {\bfseries 85} (2012) 103507}, \href{http://arxiv.org/abs/1112.1888}{{\ttfamily arXiv:1112.1888 [hep-ph]}}.

\bibitem{Huber:2013kj}
S.~J. Huber and M.~Sopena, ``{An efficient approach to electroweak bubble velocities},'' \href{http://arxiv.org/abs/1302.1044}{{\ttfamily arXiv:1302.1044 [hep-ph]}}.

\bibitem{Konstandin:2010dm}
T.~Konstandin and J.~M. No, ``{Hydrodynamic obstruction to bubble expansion},'' \href{http://dx.doi.org/10.1088/1475-7516/2011/02/008}{{\em JCAP} {\bfseries 02} (2011) 008}, \href{http://arxiv.org/abs/1011.3735}{{\ttfamily arXiv:1011.3735 [hep-ph]}}.

\bibitem{Bodeker:2009qy}
D.~Bodeker and G.~D. Moore, ``{Can electroweak bubble walls run away?},'' \href{http://dx.doi.org/10.1088/1475-7516/2009/05/009}{{\em JCAP} {\bfseries 05} (2009) 009}, \href{http://arxiv.org/abs/0903.4099}{{\ttfamily arXiv:0903.4099 [hep-ph]}}.

\bibitem{Bodeker:2017cim}
D.~Bodeker and G.~D. Moore, ``{Electroweak Bubble Wall Speed Limit},'' \href{http://dx.doi.org/10.1088/1475-7516/2017/05/025}{{\em JCAP} {\bfseries 05} (2017) 025}, \href{http://arxiv.org/abs/1703.08215}{{\ttfamily arXiv:1703.08215 [hep-ph]}}.

\bibitem{Hoche:2020ysm}
S.~H\"oche, J.~Kozaczuk, A.~J. Long, J.~Turner, and Y.~Wang, ``{Towards an all-orders calculation of the electroweak bubble wall velocity},'' \href{http://dx.doi.org/10.1088/1475-7516/2021/03/009}{{\em JCAP} {\bfseries 03} (2021) 009}, \href{http://arxiv.org/abs/2007.10343}{{\ttfamily arXiv:2007.10343 [hep-ph]}}.

\bibitem{Azatov:2020ufh}
A.~Azatov and M.~Vanvlasselaer, ``{Bubble wall velocity: heavy physics effects},'' \href{http://dx.doi.org/10.1088/1475-7516/2021/01/058}{{\em JCAP} {\bfseries 01} (2021) 058}, \href{http://arxiv.org/abs/2010.02590}{{\ttfamily arXiv:2010.02590 [hep-ph]}}.

\bibitem{Gouttenoire:2021kjv}
Y.~Gouttenoire, R.~Jinno, and F.~Sala, ``{Friction pressure on relativistic bubble walls},'' \href{http://dx.doi.org/10.1007/JHEP05(2022)004}{{\em JHEP} {\bfseries 05} (2022) 004}, \href{http://arxiv.org/abs/2112.07686}{{\ttfamily arXiv:2112.07686 [hep-ph]}}.

\bibitem{GarciaGarcia:2022yqb}
I.~Garcia~Garcia, G.~Koszegi, and R.~Petrossian-Byrne, ``{Reflections on bubble walls},'' \href{http://dx.doi.org/10.1007/JHEP09(2023)013}{{\em JHEP} {\bfseries 09} (2023) 013}, \href{http://arxiv.org/abs/2212.10572}{{\ttfamily arXiv:2212.10572 [hep-ph]}}.

\bibitem{Ai:2023suz}
W.-Y. Ai, ``{Logarithmically divergent friction on ultrarelativistic bubble walls},'' \href{http://dx.doi.org/10.1088/1475-7516/2023/10/052}{{\em JCAP} {\bfseries 10} (2023) 052}, \href{http://arxiv.org/abs/2308.10679}{{\ttfamily arXiv:2308.10679 [hep-ph]}}.

\bibitem{DeCurtis:2023hil}
S.~De~Curtis, L.~Delle~Rose, A.~Guiggiani, A.~Gil~Muyor, and G.~Panico, ``{Collision integrals for cosmological phase transitions},'' \href{http://dx.doi.org/10.1007/JHEP05(2023)194}{{\em JHEP} {\bfseries 05} (2023) 194}, \href{http://arxiv.org/abs/2303.05846}{{\ttfamily arXiv:2303.05846 [hep-ph]}}.

\bibitem{Ai:2024shx}
W.-Y. Ai, X.~Nagels, and M.~Vanvlasselaer, ``{Criterion for ultra-fast bubble walls: the impact of hydrodynamic obstruction},'' \href{http://dx.doi.org/10.1088/1475-7516/2024/03/037}{{\em JCAP} {\bfseries 03} (2024) 037}, \href{http://arxiv.org/abs/2401.05911}{{\ttfamily arXiv:2401.05911 [hep-ph]}}.

\bibitem{Leitao:2014pda}
L.~Leitao and A.~Megevand, ``{Hydrodynamics of phase transition fronts and the speed of sound in the plasma},'' \href{http://dx.doi.org/10.1016/j.nuclphysb.2014.12.008}{{\em Nucl. Phys. B} {\bfseries 891} (2015) 159--199}, \href{http://arxiv.org/abs/1410.3875}{{\ttfamily arXiv:1410.3875 [hep-ph]}}.

\bibitem{Laine:1993ey}
M.~Laine, ``{Bubble growth as a detonation},'' \href{http://dx.doi.org/10.1103/PhysRevD.49.3847}{{\em Phys. Rev. D} {\bfseries 49} (1994) 3847--3853}, \href{http://arxiv.org/abs/hep-ph/9309242}{{\ttfamily arXiv:hep-ph/9309242}}.

\bibitem{Kozaczuk:2015owa}
J.~Kozaczuk, ``{Bubble Expansion and the Viability of Singlet-Driven Electroweak Baryogenesis},'' \href{http://dx.doi.org/10.1007/JHEP10(2015)135}{{\em JHEP} {\bfseries 10} (2015) 135}, \href{http://arxiv.org/abs/1506.04741}{{\ttfamily arXiv:1506.04741 [hep-ph]}}.

\bibitem{Espinosa:1993bs}
J.~R. Espinosa and M.~Quiros, ``{The Electroweak phase transition with a singlet},'' \href{http://dx.doi.org/10.1016/0370-2693(93)91111-Y}{{\em Phys. Lett. B} {\bfseries 305} (1993) 98--105}, \href{http://arxiv.org/abs/hep-ph/9301285}{{\ttfamily arXiv:hep-ph/9301285}}.

\bibitem{Espinosa:2011ax}
J.~R. Espinosa, T.~Konstandin, and F.~Riva, ``{Strong Electroweak Phase Transitions in the Standard Model with a Singlet},'' \href{http://dx.doi.org/10.1016/j.nuclphysb.2011.09.010}{{\em Nucl. Phys. B} {\bfseries 854} (2012) 592--630}, \href{http://arxiv.org/abs/1107.5441}{{\ttfamily arXiv:1107.5441 [hep-ph]}}.

\bibitem{Niemi:2021qvp}
L.~Niemi, P.~Schicho, and T.~V.~I. Tenkanen, ``{Singlet-assisted electroweak phase transition at two loops},'' \href{http://dx.doi.org/10.1103/PhysRevD.103.115035}{{\em Phys. Rev. D} {\bfseries 103} no.~11, (2021) 115035}, \href{http://arxiv.org/abs/2103.07467}{{\ttfamily arXiv:2103.07467 [hep-ph]}}. [Erratum: Phys.Rev.D 109, 039902 (2024)].

\bibitem{Gould:2021oba}
O.~Gould and T.~V.~I. Tenkanen, ``{On the perturbative expansion at high temperature and implications for cosmological phase transitions},'' \href{http://dx.doi.org/10.1007/JHEP06(2021)069}{{\em JHEP} {\bfseries 06} (2021) 069}, \href{http://arxiv.org/abs/2104.04399}{{\ttfamily arXiv:2104.04399 [hep-ph]}}.

\bibitem{Lewicki:2024xan}
M.~Lewicki, M.~Merchand, L.~Sagunski, P.~Schicho, and D.~Schmitt, ``{Impact of theoretical uncertainties on model parameter reconstruction from GW signals sourced by cosmological phase transitions},'' \href{http://dx.doi.org/10.1103/PhysRevD.110.023538}{{\em Phys. Rev. D} {\bfseries 110} no.~2, (2024) 023538}, \href{http://arxiv.org/abs/2403.03769}{{\ttfamily arXiv:2403.03769 [hep-ph]}}.

\bibitem{Dashko:2024spj}
A.~Dashko and A.~Ekstedt, ``{Bubble-wall speed with loop corrections},'' \href{http://arxiv.org/abs/2411.05075}{{\ttfamily arXiv:2411.05075 [hep-ph]}}.

\bibitem{Ai:forthcoming}
W.-Y. Ai, M.~Carosi, B.~Garbrech, C.~Tamarit, and M.~Vanvlasselaer, ``{Bubble wall dynamics from non-equilibrium quantum field theory}.'' {\it forthcoming}.

\bibitem{Giese:2020rtr}
F.~Giese, T.~Konstandin, and J.~van~de Vis, ``{Model-independent energy budget of cosmological first-order phase transitions\textemdash{}A sound argument to go beyond the bag model},'' \href{http://dx.doi.org/10.1088/1475-7516/2020/07/057}{{\em JCAP} {\bfseries 07} no.~07, (2020) 057}, \href{http://arxiv.org/abs/2004.06995}{{\ttfamily arXiv:2004.06995 [astro-ph.CO]}}.

\bibitem{Giese:2020znk}
F.~Giese, T.~Konstandin, K.~Schmitz, and J.~van~de Vis, ``{Model-independent energy budget for LISA},'' \href{http://dx.doi.org/10.1088/1475-7516/2021/01/072}{{\em JCAP} {\bfseries 01} (2021) 072}, \href{http://arxiv.org/abs/2010.09744}{{\ttfamily arXiv:2010.09744 [astro-ph.CO]}}.

\end{thebibliography}\endgroup

\end{document}